%% file: main.tex
\pdfoutput=1

\documentclass[conference,9pt]{IEEEtran}

\usepackage{cite}
\usepackage{amsmath,amssymb,amsfonts}
\usepackage{algorithmic}
\usepackage{graphicx}
\usepackage{textcomp}
\usepackage{xcolor}

\usepackage{booktabs}
\usepackage{siunitx}
\usepackage{stfloats}
\usepackage{placeins}
\usepackage{array}
\usepackage{pifont}
\usepackage{fontawesome5}

\usepackage[hidelinks]{hyperref}

\usepackage{tabularx}
\newcolumntype{Y}{>{\raggedright\arraybackslash}X}

\usepackage{tikz}
\usetikzlibrary{arrows.meta}

\definecolor{rmsebar}{RGB}{44,125,180}
\definecolor{maebar}{RGB}{162,197,222}
\definecolor{maxaebar}{RGB}{242,145,70}

\newcommand{\benchname}{OpenTherm}
\newcommand{\scopeword}{Scope}

\newcommand{\scopeprefix}{S}
\newcommand{\scopeid}[1]{\scopeprefix#1}

\newcommand{\scoperange}[2]{\scopeid{#1}--\scopeid{#2}}

\newcommand{\topmae}{Top-50 MAE}
\newcommand{\openyes}{\textcolor{green!45!black}{\ding{51}}}
\newcommand{\openno}{\textcolor{red!70!black}{\ding{55}}}

\def\BibTeX{{\rm B\kern-.05em{\sc i\kern-.025em b}\kern-.08em
    T\kern-.1667em\lower.7ex\hbox{E}\kern-.125emX}}
\begin{document}

\title{OpenTherm: A Unified and Progressive Benchmark for 2.5D/3D-IC Thermal Prediction}

\author{%
\IEEEauthorblockN{David Hang\textsuperscript{1,*},
Wenkai Yang\textsuperscript{2,*},
Kuiye Ding\textsuperscript{1,*},
Haiyang Xin\textsuperscript{3},
Jacky Wei\textsuperscript{1}}
\IEEEauthorblockA{\textsuperscript{1}University of Technology Sydney, Sydney, NSW, Australia\\
\textsuperscript{2}School of Information Science and Technology, ShanghaiTech University, Shanghai, China\\
\textsuperscript{3}Technische Universit\"at M\"unchen, Munich, Germany\\
\textsuperscript{*}These authors contributed equally to this work.}
}

\maketitle

\begin{abstract}
Accurate and efficient chip thermal prediction is essential for advanced packaging design. In recent years, deep learning-based thermal predictors have made significant progress, achieving high prediction accuracy while substantially reducing computational costs. However, differences in datasets, simulators, and evaluation protocols hinder direct comparison of existing methods and systematic assessment of their generalization across design variations and cross-package transfer scenarios. We introduce \textbf{OpenTherm}, an open benchmark combining 32,000 established 3D-IC samples with 50,000 new 2.5D chiplet samples generated with the open-source HotSpot simulator. It provides fixed data splits, a common input representation, and shared evaluation tools. Five deployment-aligned Generalization Scopes cover fixed-design workload prediction, within-family layout, material, and cooling exploration, and cross-package out-of-distribution (OOD) transfer with few-shot adaptation. Under a shared protocol, eight model configurations show gradual error growth within known package families. The best MAE increases by $13.7\times$ from the broadest within-family scope to zero-shot cross-package transfer. Larger models do not consistently generalize better, and ten labeled fields per unseen package reduce MAE by 76--93\%. Project site: \url{https://github.com/Day333/OpenTherm}.

\end{abstract}


\section{Introduction}

Thermal analysis is essential to chip design and poses a major computational challenge for 2.5D and 3D integrated circuits (ICs), where chiplet placement, power density, packaging materials, and cooling conditions jointly influence temperature distributions and hotspot formation. Repeated high-fidelity thermal simulations during floorplanning and design-space exploration (DSE) can be computationally expensive, motivating fast learning-based surrogates. Early convolutional and generative models enabled rapid workload-conditioned thermal estimation~\cite{jin2020thermgan}. Subsequent physics-informed operator learning extended prediction to varying design configurations without labeled simulation data~\cite{liu2023deepoheat}, while recent pretrained partial differential equation (PDE) foundation models have achieved sub-0.05~K prediction errors on selected fixed-design thermal tasks~\cite{huang2026thermfm}.

Despite these advances, reported accuracies are difficult to compare across studies. Thermal predictors are evaluated with different simulators, physical configurations, spatial resolutions, input representations, splits, and metrics (Table~\ref{tab:prior-predictors-full}). Published errors are therefore not directly comparable, even for the same predictor. Open datasets and implementations help reproduce individual studies. Standardized benchmarks such as ImageNet~\cite{deng2009imagenet}, PDEBench~\cite{takamoto2022pdebench}, and CircuitNet~\cite{chai2023circuitnet} show how shared datasets, tasks, and evaluation protocols make results directly comparable.

An accurate fixed-design predictor can also face a different distribution when deployed in DSE. Designers may move chiplets during floorplanning, change filler conductivity during material selection, or vary cooling conditions within a package family whose interposer footprint, layer-stack template, and chiplet inventory stay fixed. Model reuse on a new product adds changes in chiplet counts, die sizes, and power distributions. Prior work addresses layout variation~\cite{chen2022thermgcn}, coupled changes in geometry and thermal parameters~\cite{zhang2025fsaheat}, and heterogeneous packages~\cite{lu2026cool}. However, their evaluation settings differ in the factors varied within a package family and the changes tested at transfer. A common protocol can show how the same predictor behaves across these stages.

We organize thermal-prediction evaluation around three
deployment regimes: fixed-design workload prediction,
within-family design-space exploration, and cross-package
model reuse. These regimes are instantiated as five
Generalization Scopes (Fig.~\ref{fig:benchmark-overview}).
\scopeid{1} anchors fixed-design prediction with established
steady-state, transient, and industrial 3D-IC tasks.
The HotSpot-generated \scoperange{2}{4} suite contains
45,000 samples from ten known 2.5D package families,
progressively introducing layout, filler conductivity,
and cooling-boundary variations.
\scopeid{5} evaluates frozen transfer and few-shot adaptation
on five held-out package architectures with shifts in
chiplet geometry, power distribution, and silicon utilization.
Within each generated scope, all models share fixed splits,
input representations, reference temperatures, and metrics.

The benchmark reveals a substantial gap between
within-family prediction and cross-package generalization.
Across eight model configurations, the best MAE increases
from 0.248~K in \scopeid{2} to 0.585~K in \scopeid{4},
but reaches 8.003~K under zero-shot transfer in \scopeid{5},
a 13.7$\times$ increase over the best \scopeid{4} result.
The largest Therm-FM variant leads cooling co-design,
whereas the smallest performs best in frozen cross-package transfer.
With only ten labeled fields per unseen package, few-shot
adaptation reduces MAE by 76--93\% relative to each model's
frozen performance on the same holdout.
These findings highlight the need to evaluate within-family
accuracy, zero-shot transfer, and adaptation separately.

Our contributions are as follows:
\begin{itemize}
    \item \emph{An open, progressive generalization benchmark.}
    To the best of our knowledge, OpenTherm is the first
    open dataset and benchmark jointly covering fixed-design
    prediction, within-family design exploration, and
    package-disjoint transfer through five evaluation scopes,
    with controlled target-label budgets for adaptation.

    \item \emph{An extensible suite for training and fair comparison.}
    We combine 32,000 established 3D-IC samples with
    50,000 newly generated HotSpot-based 2.5D chiplet samples,
    supported by a parametric generator, fixed splits,
    and shared model and evaluation interfaces.
    These components support reproducible training, fair
    comparison, and convenient integration of new models,
    datasets, and evaluation settings.

    \item \emph{A systematic study of thermal generalization.}
    We evaluate eight configurations using shared inputs,
    labels, splits, and metrics within each task.
    Most models retain low absolute errors under broader
    within-family variation, but degrade sharply under
    zero-shot transfer to unseen packages.
    Few-shot adaptation substantially reduces this gap, showing promise for practical cross-package generalization.
\end{itemize}

\begin{table*}[!t]
\caption{Learning-based chip and package thermal prediction.
$G$: geometry/placement; $P$: power; $M$: material;
$B$: boundary/cooling; $t$: time; NR: not reported.
Closest \scopeword{} indicates conceptual alignment.
\openyes/\openno{} indicate public availability.}
\label{tab:prior-predictors-full}
\centering
\footnotesize
\renewcommand{\arraystretch}{1.08}
\setlength{\tabcolsep}{2.2pt}

\newcommand{\surveywork}[2]{\mbox{#1~\cite{#2}}}
\newcommand{\surveyvenue}[2]{\mbox{#1'#2}}

\begin{tabularx}{\textwidth}{@{}
    >{\raggedright\arraybackslash}p{2.35cm}
    >{\raggedright\arraybackslash}p{1.75cm}
    >{\raggedright\arraybackslash}p{2.25cm}
    >{\raggedright\arraybackslash}p{2.27cm}
    >{\centering\arraybackslash}p{1.07cm}
    @{\hspace{5pt}}Y
    >{\centering\arraybackslash}p{0.85cm}
    >{\centering\arraybackslash}p{0.72cm}
    >{\centering\arraybackslash}p{0.72cm}
@{}}
\toprule
Work & Venue & Model family & Labels / scale & Factors &
Generalization setting & Closest\newline \scopeword{} &
Data\newline public & Code\newline public \\
\midrule

\surveywork{ThermGAN}{jin2020thermgan}
& \surveyvenue{ICCAD}{20}
& Conditional GAN & Commercial CPU traces; NR & $P,t$
& Workload-conditioned maps on fixed CPU designs
& \scopeid{1} & \openno & \openno \\

\surveywork{ThermEDGe}{chhabria2021thermedge}
& \surveyvenue{ASP-DAC}{21}
& \mbox{Encoder--decoder}\newline with ConvLSTM
& Private numerical; 5k & $P,t$
& Within-design steady-state and transient thermal and IR tasks
& \scopeid{1} & \openno & \openyes \\

\surveywork{GCN+PNA}{chen2022thermgcn}
& \surveyvenue{ASP-DAC}{22}
& Graph convolution & HotSpot; six unseen sets & $GP$
& Inductive evaluation on unseen chiplet layouts and sizes
& \scopeid{2}/\scopeid{5} & \openno & \openno \\

\surveywork{DeepOHeat}{liu2023deepoheat}
& \surveyvenue{DAC}{23}
& DeepONet operator & PDE residuals; NR & $GPMB$
& Physics-informed training in parametric and non-parametric setups
& \scopeid{4} & -- & \openyes \\

\surveywork{MCM-DCNN}{hua2023mcmdcnn}
& \surveyvenue{TSEP}{23}
& \mbox{SDF-encoded}\newline DCNN & Numerical; NR & $GP$
& Trains on 2--4 chips; tests on a 5-chip configuration
& \scopeid{2}/\scopeid{5} & \openno & \openno \\

\surveywork{ARO}{wang2024aro}
& \surveyvenue{ICCAD}{24}
& Autoregressive\newline operator
& MTA~\cite{ladenheim2018mta}; $3\times$5k & $P,t$
& Within-design power variation; cross-case transfer and multi-fidelity analysis
& \scopeid{1} & \openyes & \openyes \\

\surveywork{FaStTherm}{zhu2024fasttherm}
& \surveyvenue{ICCAD}{24}
& Residual CNN & Full-chip transient; NR & $P,t$
& Nonlinear transient prediction on task-specific designs
& \scopeid{1} & \openno & \openno \\

\surveywork{T-Fusion}{zhang2025tfusion}
& \surveyvenue{ASP-DAC}{25}
& Bayesian\newline tensor fusion & EV6 MTA; HF/LF subsets & $P,t$
& Multi-fidelity transient prediction on a fixed design
& \scopeid{1} & \openno & \openno \\

\surveywork{SAU-FNO}{huang2025saufno}
& \surveyvenue{DAC}{25}
& Attention \mbox{U-FNO} & MTA; $3\times$5k & $P$
& Fixed-design power variation; transfer between source cases
& \scopeid{1} & \openno & \openno \\

\surveywork{FSA-Heat}{zhang2025fsaheat}
& \surveyvenue{DATE}{25}
& \mbox{Frequency-spatial}\newline net & HotSpot; 6k & $GPMB$
& Represented mixtures plus unseen conductivity and source counts
& \scopeid{3}/\scopeid{5} & \openno & \openno \\

\surveywork{Therunet-SP}{lee2025therunet}
& \surveyvenue{ISQED}{25}
& Scaled \mbox{U-Net} & 2.5D numerical; NR & $GP$
& Geometry scaling across 2.5D configurations
& \scopeid{2} & \openno & \openno \\

\surveywork{Therm-PCT}{zhang2025thermpct}
& \surveyvenue{ICCAD}{25}
& \mbox{Point-cloud}\newline Transformer & COMSOL point clouds; 8.5k & $GPM$
& Unseen shapes and point-cloud sizes on unstructured domains
& \scopeid{3}/\scopeid{5} & \openno & \openno \\

Adaptive\newline \surveywork{Graph}{yang2025adaptivegraph}
& \surveyvenue{ICCAD}{25}
& \mbox{GNN--FEM} hybrid
& Fine-grid numerical\newline + silicon
& $GM$
& Variable structures and interfaces; unseen processes and materials
& \scopeid{3}/\scopeid{5} & \openno & \openno \\

\surveywork{PNO-Therm}{huang2026pnotherm}
& \surveyvenue{DAC}{26}
& PDE foundation\newline model & HS source suite; $3\times$5k & $P$
& Fine-tuning on fixed-design steady-state tasks
& \scopeid{1} & \openno & \openno \\

\surveywork{COOL}{lu2026cool}
& \surveyvenue{DAC}{26}
& 3D point\newline Transformer & Commercial FEM; $74\times10$ & $GPMB$
& Design-disjoint 58/16 split across heterogeneous 3D/3.5D systems
& \scopeid{4}/\scopeid{5} & \openno & \openno \\

\bottomrule
\end{tabularx}
\end{table*}

\section{Related Work}
\label{sec:related-work}

\subsection{Standardized Benchmarks in ML and EDA}

Benchmarks such as ImageNet~\cite{deng2009imagenet}, GLUE~\cite{wang2019glue}, and PDEBench~\cite{takamoto2022pdebench} established shared tasks, splits, and evaluation protocols in vision, language, and scientific computing. In EDA, CircuitNet~\cite{chai2023circuitnet}, OpenABC-D~\cite{chowdhury2021openabcd}, AMSNet~\cite{tao2024amsnet}, EDALearn~\cite{pan2024edalearn}, and ChiPBench~\cite{wang2025chipbench} provide evaluation infrastructure for tasks including cross-stage physical design prediction, logic synthesis, and placement. For chip thermal prediction, Coenen et al.~\cite{coenen2023benchmarkthermal} benchmarked neural-network and reduced-order approaches using finite-element data for multiscale thermal simulation. OpenTherm evaluates modern thermal surrogates under within-family design variation and package-disjoint transfer, with a common simulator, spatial resolution, input representation, and data split for all predictors within each task.

\subsection{Learning-Based Thermal Prediction}

Early thermal surrogates used image-to-image mapping for steady-state and transient prediction on fixed layouts, including conditional GANs~\cite{jin2020thermgan}, convolutional encoder--decoders~\cite{chhabria2021thermedge,lin2023physicsunet}, and residual CNNs~\cite{zhu2024fasttherm}. FNO~\cite{li2021fno}, U-FNO~\cite{wen2022ufno}, and attention-augmented SAU-FNO~\cite{huang2025saufno} model spatial heat propagation with spectral operators. DeepOHeat~\cite{liu2023deepoheat} uses physics-informed operator learning, while ARO~\cite{wang2024aro} uses an autoregressive operator for transient prediction. PNO-Therm~\cite{huang2026pnotherm} and Therm-FM~\cite{huang2026thermfm} adapt pretrained PDE models to fixed-design thermal tasks. Training spans physics-informed residual minimization, simulated-field regression, and transfer from PDE pretraining. OpenTherm records label access and training recipes alongside architecture to contextualize comparisons.

For chiplet DSE, Therunet-SP~\cite{lee2025therunet} and MCM-DCNN~\cite{hua2023mcmdcnn} explore layout or chip-count changes, while FSA-Heat~\cite{zhang2025fsaheat} varies placement, conductivity, and convection. Graph-based predictors~\cite{chen2022thermgcn,yang2025adaptivegraph} and point-cloud models such as Therm-PCT~\cite{zhang2025thermpct} and COOL~\cite{lu2026cool} address irregular geometries or heterogeneous packages; COOL uses a design-disjoint split. Table~\ref{tab:prior-predictors-full} maps these evaluations to the five scopes. Differences in simulators, representations, and held-out conditions prevent direct comparison of published errors. OpenTherm provides shared tasks and evaluation for within-family DSE, frozen package-disjoint transfer, and budgeted adaptation.

\begin{figure*}[!t]
    \centering
    \begin{tikzpicture}[
        x=.01\linewidth, y=1mm, font=\footnotesize,
        card/.style={rounded corners=1.5pt, line width=.35pt},
        title/.style={font=\footnotesize\bfseries, inner sep=1pt},
        flow/.style={draw=black!55, fill=black!3, rounded corners=1.5pt,
                     line width=.35pt, align=center, inner sep=1.3pt},
        line/.style={draw=black!70, line width=.45pt},
        arrow/.style={line, -{Latex[length=1.25mm]}}
    ]
    \path[use as bounding box] (0,0) rectangle (100,47.8);

    \node[title,anchor=west] at (.7,45.8) {(a) Scope progression};
    \draw[black!60] (20.7,42.8) -- (20.7,44.3)
        -- (79.3,44.3) -- (79.3,42.8);
    \node[fill=white,inner sep=1pt] at (50,44.3)
        {Within-family DSE (Cases 1--10)};

    \foreach \cardX/\cardID/\cardTitle/\cardDetail/\cardN/\cardColor in {
        10/S1/{Fixed design}/{Power and time}/32k/blue,
        30/S2/{Layout}/{Placement and rotation}/15k/teal,
        50/S3/{Material}/{Filler conductivity $k$}/15k/teal,
        70/S4/{Cooling}/{Ambient \& convection}/15k/teal,
        90/S5/{Unseen packages}/{Cases 11--15}/5k/orange} {
        \begin{scope}[shift={(\cardX,0)}]
            \draw[card, draw=\cardColor!70!black, fill=\cardColor!6]
                (-9.3,20) rectangle (9.3,42);
            \node[title] at (0,39.4) {\cardID{} \cardTitle};
            \node at (0,25.9) {\cardDetail};
            \node at (0,22.7) {\textbf{\cardN} samples};
        \end{scope}
    }

    \foreach \layerY in {29.4,31.3,33.2} {
        \draw[draw=blue!65!black, fill=blue!22, line width=.35pt]
            (5.7,\layerY) rectangle ++(8.6,1.35);
    }

    \foreach \cardX/\baseFill in {30/white,50/teal!18,70/teal!18} {
        \begin{scope}[shift={(\cardX,0)}]
            \draw[draw=teal!60!black, fill=\baseFill, line width=.35pt]
                (-5.5,29.3) rectangle (5.5,35.2);
            \fill[teal!65!black] (-4.3,30.1) rectangle (-1.5,32.6);
            \fill[teal!65!black] (-.5,32.0) rectangle (2.3,34.5);
            \fill[teal!65!black] (2.8,30.2) rectangle (4.7,32.1);
        \end{scope}
    }
    \draw[arrow] (26.4,33.2) -- (27.9,34.5);
    \draw[arrow] (68.1,35.4) -- (68.1,37.3);
    \draw[arrow] (71.9,35.4) -- (71.9,37.3);

    \begin{scope}[shift={(90,0)}]
        \draw[draw=orange!75!black, fill=white, line width=.35pt]
            (-5.5,29.3) rectangle (5.5,35.2);
        \foreach \chipX in {-4.8,-3.05,-1.30,.45,2.20,3.95} {
            \foreach \chipY in {29.9,31.6,33.3} {
                \fill[orange!85!black]
                    (\chipX,\chipY) rectangle ++(1.1,1.15);
            }
        }
    \end{scope}

    \node[title,anchor=west] at (.7,17.3)
        {(b) Dataset generation (S2--S5)};
    \node[title,anchor=west] at (51,17.3)
        {(c) Cross-package evaluation (S5)};
    \draw[black!35] (49.6,.6) -- (49.6,17.4);

    \node[flow,text width=.146\linewidth] (gen) at (8.7,7.5)
        {Parametric layout\\and physics};
    \node[flow,text width=.207\linewidth] (input) at (35.3,11.5)
        {Rasterized input $X$\\$64\times64$ (3/4/7/7 channels)};
    \node[flow,text width=.207\linewidth] (target) at (35.3,3.5)
        {HotSpot simulation\\Temperature label $Y$ (K)};

    \draw[line] (gen.east) -- ++(2.2,0) coordinate (genfork);
    \draw[arrow] (genfork) |- (input.west);
    \draw[arrow] (genfork) |- (target.west);

     \node[flow,text width=.136\linewidth] (ckpt) at (58.8,7.5)
        {S4-trained\\source model};

    \node[flow,text width=.221\linewidth] (zero) at (86.8,11.5)
        {Frozen zero-shot\\S5 evaluation};

    \node[flow,text width=.221\linewidth] (adapt) at (86.8,3.5)
        {$K$-shot adaptation\\Held-out S5 evaluation};

    \draw[line] (ckpt.east) -- ++(2.2,0) coordinate (evalfork);
    \draw[arrow] (evalfork) |- (zero.west);
    \draw[arrow] (evalfork) |- (adapt.west);
    \end{tikzpicture}
    
\caption{Overview of OpenTherm. Five progressive
Generalization Scopes (\scoperange{1}{5}) align thermal
evaluation with practical EDA deployment stages, covering
fixed-design prediction (\scopeid{1}), within-family
design-space exploration (\scoperange{2}{4}), and
cross-package generalization (\scopeid{5}). The S5
evaluation includes frozen zero-shot transfer and
few-shot adaptation from S4-trained checkpoints.}
\label{fig:benchmark-overview}
\end{figure*}

\section{Problem Formulation}
\label{sec:problem}
Within an IC package, the temperature field $T(\mathbf{r},t)$ obeys the heat conduction equation
\begin{equation}
\rho(\mathbf{r}) C_p(\mathbf{r}) \frac{\partial T}{\partial t}
- \nabla\cdot\bigl(k(\mathbf{r})\nabla T\bigr) = Q(\mathbf{r},t),
\label{eq:heat}
\end{equation}
where $\rho$, $C_p$, $k$, and $Q$ denote density, specific heat capacity, thermal conductivity, and volumetric heat-generation rate, respectively. At a convective boundary, heat exchange is described by
\begin{equation}
-\mathbf{n}\cdot\bigl(k\nabla T\bigr)=h(T-T_{\mathrm{amb}}),
\label{eq:cooling-boundary}
\end{equation}
where $T_{\mathrm{amb}}$ is the ambient temperature, $h$ is the convection heat transfer coefficient, and $\mathbf{n}$ is the outward unit normal. Temperature and normal heat flux are continuous across perfectly bonded material interfaces. The time derivative vanishes at steady state; transient tasks additionally specify an initial temperature field. For the generated \scoperange{2}{5} suite, reference temperatures are computed using HotSpot's equivalent thermal network, with an effective heatsink-to-ambient resistance representing package cooling.

We learn a parametric surrogate of the thermal solution operator,
\begin{equation}
\mathcal{G}_{\theta}\colon (\mathcal{M},\mathbf{z})\mapsto \widehat{T},
\label{eq:thermal-operator}
\end{equation}
where $\mathcal{M}$ denotes the package family. In \scoperange{2}{5}, each family fixes the interposer footprint, layer-stack template, and chiplet inventory, including die dimensions. The task variables $\mathbf{z}$ describe placement, power, material conductivity, cooling, and time where applicable. Each sample is represented by a gridded input--target pair $(\mathbf{X}_j,\mathbf{Y}_j)$, and the predictor produces $\widehat{\mathbf{Y}}_j=f_\theta(\mathbf{X}_j)$. Within each task, the benchmark specifies a common input schema, reference labels, and evaluation split.

Within each \scopeid{1} task, training and test samples share the same design and vary in workload and, for transient prediction, time. In \scoperange{2}{4}, the same ten package families appear in training and testing, and successive scopes expand the physical factors sampled within those families. For zero-shot \scopeid{5}, the source and test family sets are disjoint, $\mathfrak{M}_{\mathrm{train}}\cap\mathfrak{M}_{\mathrm{test}}=\varnothing$. Evaluating within-family variation and cross-package shift separately distinguishes generalization within known families from transfer to new packages.

\begin{table}[!t]
\caption{Summary of OpenTherm Generalization Scopes. Channels expand cumulatively from \scopeid{2} to \scopeid{4}, while \scopeid{5} retains the 7-channel \scopeid{4} schema on five unseen package architectures.}
\label{tab:benchmark-scopes}
\centering
\footnotesize
\renewcommand{\arraystretch}{1.06}
\setlength{\tabcolsep}{2.8pt}
\begin{tabularx}{\columnwidth}{@{}clccYY@{}}
\toprule
Scope & Systems & Chan. & Samples & Active Variations & Target EDA Stage \\
\midrule
\scopeid{1} & EV6 + IND & 4/5\textsuperscript{a} & 32k
& Power, time $t$ & Runtime workload \\
\scopeid{2} & Cases 1--10 & 3 & 15k
& Placement, orient. & Floorplan DSE \\
\scopeid{3} & Cases 1--10 & 4 & 15k
& \scopeid{2} + filler $k$ & Material sweep \\
\scopeid{4} & Cases 1--10 & 7 & 15k
& \scopeid{3} + $T_{\mathrm{amb}}, h, R_{\mathrm{conv}}$
& Cooling co-design \\
\scopeid{5} & Cases 11--15 & 7 & 5k
& Cross-package OOD & Cross-product reuse \\
\bottomrule
\end{tabularx}

\par\vspace{2pt}
\begin{minipage}{\columnwidth}
\scriptsize
\raggedright
\textsuperscript{a}S1 uses four physical feature channels
for steady-state inputs and five for transient inputs;
layer and temporal axes are counted separately.
\end{minipage}
\end{table}

\section{Benchmark Design and Protocol}
\label{sec:benchmark}

\subsection{Five Progressive Generalization Scopes}
\label{sec:five-scopes}

Table~\ref{tab:benchmark-scopes} summarizes the five scopes.
\scopeid{1} retains 32,000 fixed-design samples from the
Therm-FM suite~\cite{huang2026thermfm}, covering Alpha EV6
steady-state and transient prediction and industrial package tasks
(Appendix~\ref{app:level1}).

The 45,000 steady-state samples in \scoperange{2}{4} cover
ten package families (Cases~1--10) adapted from
ATPlace2.5D~\cite{wang2024atplace}. These families contain
6--61 chiplets, with interposer dimensions ranging from
$26\times23$ to $59\times61$~mm$^2$. On a $64\times64$ grid,
the resulting cell pitch ranges from approximately
0.36 to 0.95~mm across the two axes.
\scopeid{2} varies placement and orientation for floorplan
exploration; \scopeid{3} additionally varies filler conductivity
for material exploration; and \scopeid{4} further varies
ambient temperature and convection for cooling co-design.
Each scope contains 15,000 independently sampled layouts
and physical settings.

\scopeid{5} evaluates cross-package transfer on 1,000 samples
from each of five packages held out from source training
and validation. Case~11 has 80 chiplets.
Case~12 reaches 0.85~W/mm$^2$ in package power density
(training maximum: 0.745~W/mm$^2$).
Case~13 combines one $42\times16$~mm$^2$ die with
24 micro-chiplets with lateral dimensions of 2--2.6~mm.
Case~14 concentrates 90.9\% of package power in one chiplet
(training maximum: 69.2\%).
Case~15 reaches 68.1\% silicon utilization
(training maximum: 67.1\%).
Each held-out package combines multiple structural
and thermal changes (Table~\ref{tab:l5-ood}).

\subsection{Open-Source Generation and Grid Schema}
\label{sec:data-generation}

We generate all 50,000 \scoperange{2}{5} samples using HotSpot in detailed-3D steady-state grid mode~\cite{huang2006hotspot}. HotSpot is an open-source thermal simulator also used in chiplet thermal learning~\cite{chen2022thermgcn,zhang2025fsaheat}. The six-layer stack comprises, from bottom to top, substrate, epoxy/C4, silicon interposer, microbump underfill, chiplet silicon/filler, and thermal interface material (TIM). Their thicknesses are 200, 70, 110, 10, 150, and 20~$\mu$m, respectively. The corresponding thermal conductivities are 0.30, 1.6, 100, 1.6, 100 (silicon) or 0.2--5.0 (filler), and 4.0~W/(m$\cdot$K), respectively (Table~\ref{tab:appendix-stack}). The target $\mathbf{Y}$ is the $64\times64$ steady-state temperature map of the chiplet layer, in kelvin.

We generate layouts in random, dispersed, clustered, and boundary-aligned modes in a $6{:}2{:}1{:}1$ ratio (Appendix~\ref{app:data-specification}). Each layout must fit within the interposer without die overlap and preserve the chiplet dimensions and nominal powers specified by its case.

Rasterizing accepted layouts and encoding their physical settings yields the following inputs:
\begin{equation}
\begin{aligned}
\mathbf{X}_{\mathrm{S2}}
  &= [P,x,y],\\
\mathbf{X}_{\mathrm{S3}}
  &= [P,x,y,k],\\
\mathbf{X}_{\mathrm{S4/S5}}
  &= [P,x,y,k,T_{\mathrm{amb}},h,R_{\mathrm{conv}}].
\end{aligned}
\label{eq:channels}
\end{equation}
Channel $P$ assigns each occupied cell the nominal total power of its chiplet, in watts, and assigns zero to background cells. This is a broadcast chiplet-level feature, not per-cell power or power density. The resulting channel values, including background zeros, range from 0--300~W for \scoperange{2}{4} and 0--280~W for \scopeid{5}. The $x,y$ channels encode physical coordinates in millimeters, allowing predictors to distinguish physical distances across interposers represented on the same grid size.

In \scoperange{3}{5}, channel $k$ assigns 100~W/(m$\cdot$K) to silicon and the sampled filler conductivity to background cells. The filler conductivity is sampled log-uniformly from $[0.2,5.0]$~W/(m$\cdot$K). In \scopeid{4} and \scopeid{5}, ambient temperature $T_{\mathrm{amb}}$ is sampled uniformly from $[25,55]\,^{\circ}$C, and the convection heat-transfer coefficient $h$ is sampled log-uniformly from $[500,20000]$~W/(m$^2$K). The effective heatsink-to-ambient thermal resistance is derived as $R_{\mathrm{conv}}=1/(hA_{\mathrm{sink}})$, where $A_{\mathrm{sink}}$ is the heatsink area used in the cooling model, in m$^2$. This area is determined by the package dimensions, so $R_{\mathrm{conv}}$ also encodes case-dependent cooling geometry. All three cooling quantities are broadcast over the grid, with $T_{\mathrm{amb}}$ stored in kelvin. The generator additionally retains chiplet identities and geometry as layout metadata; predictors receive the channels listed in Eq.~\eqref{eq:channels}. Channel semantics and empirical input ranges are listed in Tables~\ref{tab:model-input-channels} and~\ref{tab:appendix-ranges}.

\scopeid{1} retains the source tasks' reference labels and fidelity settings. For \scoperange{2}{5}, all models use the same HotSpot-generated labels, and the open-source solver and generator support local reproduction and extension.

\begin{table}[!t]
\caption{Baseline configurations, parameter counts, and training costs for \scopeid{4}. Costs cover benchmark training on one RTX A6000 and exclude pretraining and data generation.}
\label{tab:baseline-configs}
\centering
\footnotesize
\renewcommand{\arraystretch}{1.04}
\setlength{\tabcolsep}{3pt}
\begin{tabularx}{\columnwidth}{@{}lYrr@{}}
\toprule
Model & Architecture & Params & GPU$\cdot$s \\
\midrule
U-Net & 58/116/232/348 channels, 3 downsampling stages & 5.06M & 875 \\
FNO & FNO3d, modes $(12,12,1)$, width 72 & 11.98M & 1,867 \\
U-FNO & Net3d, modes $(10,10,1)$, width 36 & 5.10M & 11,658 \\
SAU-FNO & U-FNO with self-attention, same modes and width & 5.11M & 22,300 \\
DeepOHeat & 128-D trunk, 256-wide 7-layer branch & 7.92M & \textbf{264} \\
Therm-FM-T & Pretrained scOT/Poseidon & 20.7M & 3,297 \\
Therm-FM-B & Pretrained scOT/Poseidon & 157.6M & 6,580 \\
Therm-FM-L & Pretrained scOT/Poseidon & 628.4M & 11,030 \\
\bottomrule
\end{tabularx}
\end{table}

\subsection{Reproducible Interface and Extension}
\label{sec:benchmark-interface}

The toolkit separates model implementation from data handling and evaluation, making new predictors and evaluation settings straightforward to integrate. It connects generation, tensor conversion, training, inference, and scoring through reusable components. For \scoperange{2}{5}, a common tensor interface accepts gridded inputs with scope-dependent channels and returns one temperature field per sample. A standard PyTorch predictor is integrated through a model implementation, a registry entry specifying its constructor and optimization recipe, and lightweight command wrappers (Appendix~\ref{app:training-details}).

The shared runner provides fixed splits, training-only normalization, six evaluation metrics, and the \scopeid{5} adaptation loop. New evaluation settings can reuse the loading, normalization, inference, and scoring components while adapting case selection, split construction, or target-label budgets. This reduces the need to rebuild evaluation pipelines when testing additional packages or generalization conditions. Saved model artifacts retain normalization information for reproducible inference. Custom training stacks use adapters to return predictions in the common layout and follow the same evaluation protocol.

Compatible datasets reuse the loader through the released input/output tensor schema. Extensions must document package identities, channel semantics and units, label provenance, and fixed splits, together with any additional training data, pretrained weights, or target-label access. New simulators, temporal outputs, or alternative representations may require additional adapters. Modified evaluation settings are reported separately from the reference tracks, supporting convenient extension while preserving a stable basis for comparison. Appendices~\ref{app:fair-comparison} and~\ref{app:extension} specify the minimum reporting record and the extension protocol.

\begin{table}[!t]
\caption{Selected predictors on the \scopeid{1} fixed-design
anchor tasks at each task's finest available spatial resolution.
Entries summarize task-level MAEs (K) within each group;
ranges denote the minimum and maximum.
Therm-FM-L results are taken from the source
study~\cite{huang2026thermfm};
the remaining rows report benchmark reruns.
Qualitative \scopeid{1} results are shown in Appendix~\ref{app:level1}.}
\label{tab:s1-summary}
\centering
\footnotesize
\renewcommand{\arraystretch}{1.05}
\setlength{\tabcolsep}{4.5pt}
\begin{tabular}{@{}lccc@{}}
\toprule
Method & EV6 steady & EV6 transient & Industrial \\
\midrule
FNO
  & 0.063--0.164 & 0.034--0.163 & 0.017--0.018 \\
U-FNO
  & 0.046--0.126 & 0.011--0.144 & 0.013 \\
SAU-FNO
  & 0.041--0.162 & 0.013--0.149 & 0.020--0.026 \\
DeepOHeat
  & 0.745--2.585 & 0.315--1.820 & 0.039--0.044 \\
Therm-FM-L (629M)
  & 0.012--0.049 & 0.004--0.030 & 0.008 \\
\bottomrule
\end{tabular}
\end{table}

\begin{table*}[!t]
\caption{Main OpenTherm results across Within-Family DSE (\scoperange{2}{4}) and Zero-Shot Cross-Package OOD (\scopeid{5}). MaxAE denotes mean per-sample worst-cell error; Top50 denotes Top-50 hotspot MAE. All values are in kelvin, with the best per column in bold. Complete six-metric results, including $R^2$ and $E_{T_{\max}}$, are given in Appendices~\ref{app:insupport-results} and~\ref{app:ood-results}.}
\label{tab:progressive-results}
\centering

\footnotesize
\renewcommand{\arraystretch}{1.12}
\setlength{\tabcolsep}{2pt}
\begin{tabular*}{\textwidth}{@{\extracolsep{\fill}}
    l@{\hspace{6pt}}
    *{4}{r}@{\hspace{6pt}}
    *{4}{r}@{\hspace{6pt}}
    *{4}{r}@{\hspace{10pt}}
    *{4}{r}@{}}
\toprule
& \multicolumn{12}{c}{\textbf{Within-Family DSE (Cases 1--10)}}
& \multicolumn{4}{c@{}}{\textbf{OOD (Cases 11--15)}} \\
\cmidrule(lr){2-13}\cmidrule(l){14-17}
& \multicolumn{4}{c}{\textbf{\scopeid{2}: Layout DSE}}
& \multicolumn{4}{c}{\textbf{\scopeid{3}: Material Sweep}}
& \multicolumn{4}{c}{\textbf{\scopeid{4}: Cooling Co-Design}}
& \multicolumn{4}{c@{}}{\textbf{\scopeid{5}: Zero-shot}} \\
\cmidrule(lr){2-5}\cmidrule(lr){6-9}
\cmidrule(lr){10-13}\cmidrule(l){14-17}
Method
& RMSE & MAE & MaxAE & Top50
& RMSE & MAE & MaxAE & Top50
& RMSE & MAE & MaxAE & Top50
& RMSE & MAE & MaxAE & Top50 \\
\midrule

FNO & 0.828 & 0.621 & 3.965 & 0.494 & 0.962 & 0.720 & 4.472 & 0.506 & 1.291 & 0.994 & 5.436 & 0.693 & 42.175 & 39.358 & 65.918 & 39.713 \\
U-Net & 0.738 & 0.541 & 5.273 & 0.500 & 0.843 & 0.630 & 5.020 & 0.502 & 1.271 & 0.976 & 7.017 & 0.838 & 18.854 & 17.374 & 32.432 & 18.064 \\
U-FNO & 0.444 & 0.326 & 2.757 & 0.298 & \textbf{0.670} & \textbf{0.493} & \textbf{3.367} & \textbf{0.373} & 0.939 & 0.724 & 5.011 & 0.573 & 28.154 & 27.060 & 39.489 & 26.647 \\
SAU-FNO & 0.534 & 0.409 & 2.632 & 0.428 & 0.708 & 0.525 & 3.700 & 0.447 & 0.953 & 0.739 & 4.137 & 0.683 & 26.121 & 25.056 & 36.358 & 26.902 \\
DeepOHeat & 3.706 & 2.938 & 10.261 & 3.645 & 3.970 & 3.226 & 10.553 & 4.826 & 4.739 & 3.903 & 11.755 & 5.924 & 23.973 & 23.141 & 33.433 & 24.078 \\
Therm-FM-T (21M) & 0.575 & 0.410 & 3.121 & 0.399 & 0.911 & 0.679 & 4.268 & 0.551 & 1.050 & 0.794 & 4.843 & 0.643 & \textbf{8.673} & \textbf{8.003} & \textbf{15.535} & \textbf{9.045} \\
Therm-FM-B (158M) & 0.461 & 0.320 & 2.709 & 0.311 & 0.722 & 0.517 & 3.717 & 0.378 & 0.846 & 0.618 & 4.260 & 0.461 & 13.055 & 12.339 & 21.286 & 15.746 \\
Therm-FM-L (629M) & \textbf{0.354} & \textbf{0.248} & \textbf{2.145} & \textbf{0.238} & 0.737 & 0.527 & 3.785 & 0.390 & \textbf{0.811} & \textbf{0.585} & \textbf{4.034} & \textbf{0.431} & 15.336 & 14.633 & 24.088 & 13.768 \\
\bottomrule
\end{tabular*}\par
\vspace*{-3mm}
\end{table*}

\section{Experimental Evaluation}
\label{sec:experiments}

\subsection{Evaluation Protocol and Metrics}
\label{sec:evaluation-protocol}

In each of \scoperange{2}{4}, we use fixed splits of
1,080 training, 120 validation, and 300 test samples
per case, yielding 10,800/1,200/3,000 samples per scope.
Each of the ten cases is equally represented in every split.
The scopes are sampled independently and have separate
test sets of equal size.
Input normalization is fitted per channel on the training
split, and target normalization is fitted on the training
labels. Both remain frozen for validation and testing.

The official \scopeid{5} zero-shot track evaluates frozen
\scopeid{4} checkpoints on all 5,000 target samples.
Cases~11--15 are excluded from source training, validation,
checkpoint selection, and normalization fitting.
For the adaptation track, the first 500 released samples
per case form the label pool, and the final 500 form a
fixed test holdout.
For $K\in\{10,50,100,250,500\}$, the first $K$ samples
from each pool define fixed nested subsets.
Each predictor is jointly fine-tuned on the combined
$5K$ labeled fields.
Each run starts independently from the corresponding
\scopeid{4} checkpoint, with a fresh optimizer and frozen
source normalization.
We report results from the checkpoint after 50
fine-tuning epochs.
Learning rates are $10^{-4}$ for non-Therm-FM models
and $1.5\times10^{-5}$ for Therm-FM.
The holdout is used only for evaluation and does not
contribute to gradient updates or checkpoint selection.
At $K=0$, the frozen source model is evaluated on this
same 2,500-sample holdout, providing a directly comparable
baseline for adaptation.
Appendix~\ref{app:evaluation-contract} details the splits,
OOD protocol, and metric definitions.

Metrics are computed after inverse normalization.
For a single-layer field in sample $j$ with $N$ evaluated
cells, let $d_{j,i}=\widehat{T}_{j,i}-T_{j,i}$,
and let $H_j$ contain the 50 hottest cells in the reference
field. We use
\begin{equation}
\begin{aligned}
\mathrm{MAE}_j
  &= N^{-1}\textstyle\sum_i |d_{j,i}|,\\
\mathrm{RMSE}_j
  &= \sqrt{N^{-1}\textstyle\sum_i d_{j,i}^{2}},\\
\mathrm{MaxAE}_j
  &= \max_i |d_{j,i}|,\\
E_{T_{\max},j}
  &= \left|\max_i\widehat{T}_{j,i}
       -\max_iT_{j,i}\right|,\\
\mathrm{Top50}_j
  &= 50^{-1}\textstyle\sum_{i\in H_j}|d_{j,i}|.
\end{aligned}
\label{eq:error-metrics}
\end{equation}
For multi-layer \scopeid{1} fields, these metrics are
computed separately for each layer and averaged within
each sample. The resulting sample-level errors are then
averaged over the test set and reported in kelvin.
Thus, MaxAE is the mean per-sample worst-cell error for
single-layer tasks and the mean of layerwise worst-cell
errors for multi-layer tasks.
Peak-temperature error compares maxima that may occur
at different locations, whereas Top-50 MAE evaluates
predictions at reference hotspot locations.

We compute $R^2$ separately for each output channel by
pooling all test samples and cells, using that channel's
pooled reference mean in the denominator.
The resulting dimensionless scores are averaged over
output channels.
Together, these metrics assess field reconstruction,
worst-cell errors, and hotspot-temperature accuracy.

\subsection{Baselines and Training}
\label{sec:experimental-setup}

We evaluate eight configurations: a convolutional
U-Net~\cite{ronneberger2015unet}; three Fourier operators,
FNO, U-FNO, and SAU-FNO~\cite{li2021fno,wen2022ufno,huang2025saufno};
DeepOHeat~\cite{liu2023deepoheat}; and three pretrained
Therm-FM variants~\cite{huang2026thermfm}
(Table~\ref{tab:baseline-configs}; full configurations in
Appendix~\ref{app:training-details}).
Our DeepOHeat baseline retains the branch--trunk architecture
but uses supervised field regression with the same simulator
labels as the other models.
For methods without complete public training code, we use
reference implementations based on the published descriptions.

For \scoperange{2}{4}, each configuration is trained for
100 epochs on the corresponding scope.
Non-Therm-FM models minimize MSE on normalized temperature
fields using Adam with a learning rate of $10^{-3}$,
weight decay of $10^{-4}$, and batch size 20.
StepLR reduces the learning rate by a factor of 0.9
every two epochs, and evaluation uses the final checkpoint.

Therm-FM uses AdamW with a loss that divides MSE by the
mean squared target value, with both terms computed
after target normalization.
The base learning rate is $1.5\times10^{-4}$, while the
embedding and recovery modules use $1.5\times10^{-3}$.
We use weight decay of $10^{-6}$, batch size 40,
and a cosine learning-rate schedule without warmup.
The checkpoint with the lowest validation loss is selected.
Table~\ref{tab:appendix-optimization} summarizes both training recipes.
The separate \scopeid{5} adaptation settings are specified
in Section~\ref{sec:evaluation-protocol}.

All \scoperange{2}{5} experiments use a single
NVIDIA RTX A6000 GPU (48~GB), PyTorch~2.0.1, and CUDA~11.8.

\subsection{Fixed-Design Anchor and Within-Family DSE (\scoperange{1}{4})}
\label{sec:s1-s4-results}

Table~\ref{tab:s1-summary} summarizes task-level MAE ranges
within the fixed-design task groups.
FNO, U-FNO, and SAU-FNO achieve MAEs below 0.17~K
on all reported \scopeid{1} tasks.
The source-study results for Therm-FM-L span
0.004--0.049~K.
These results provide a fixed-design accuracy reference.
We next assess within-family DSE on the common HotSpot
suite in \scoperange{2}{4}.

On the within-family track, the lowest reported MAE
increases from 0.248~K in \scopeid{2} to 0.493~K in
\scopeid{3} and 0.585~K in \scopeid{4}
(Table~\ref{tab:progressive-results}; complete results in
Appendix~\ref{app:insupport-results}).
All eight configurations show increasing MAE across
these scopes.
Because the scopes are sampled independently, this trend
characterizes the complete tasks without isolating the
effect of any individual design variable.
In \scopeid{4}, all models except DeepOHeat retain
test-set MAEs below 1~K.

U-FNO achieves the lowest values on all four metrics
reported for \scopeid{3}, including 0.493~K MAE
versus 0.527~K for Therm-FM-L.
Therm-FM-L leads the same four metrics in
\scopeid{2} and \scopeid{4}.
With approximately 5.1M parameters, U-FNO remains
within $1.26\times$ of the lowest reported RMSE
in each of the three within-family scopes,
providing a compact baseline with competitive
field-reconstruction accuracy.

Model rankings also depend on the error metric.
In \scopeid{4}, Therm-FM-B has lower MAE than SAU-FNO
(0.618 vs.\ 0.739~K), whereas SAU-FNO has lower MaxAE
(4.137 vs.\ 4.260~K).
Thus, lower average field error does not necessarily
imply a lower mean per-sample worst-cell error.

\begin{figure*}[!t]
    \centering
    \begin{tikzpicture}[
        x=.01\linewidth, y=1mm, font=\scriptsize,
        axis/.style={black!65,line width=.34pt},
        guide/.style={black!18,line width=.25pt},
        bar/.style={line width=.18pt},
        model/.style={anchor=north,inner sep=0pt},
        charttitle/.style={font=\footnotesize,inner sep=0pt}]
    \foreach \x/\shade/\metric in {35/rmsebar/RMSE,46/maebar/MAE,56/maxaebar/MaxAE}{
        \filldraw[bar,fill=\shade,draw=\shade!75!black] (\x,29.1) rectangle ++(1.1,1.55);
        \node[anchor=west,inner sep=1.4pt] at ({\x+1.1},29.9) {\metric};
    }
    \node[rotate=90,inner sep=0pt] at (0.8,14.5) {Error (K)};
    \begin{scope}[shift={(6,5.5)}]
        \node[charttitle] at (21.75,21.8) {(a) Frozen zero-shot ($K=0$)};
        \foreach \tick in {0,20,40,60}{
            \pgfmathsetmacro{\gy}{19*\tick/70}
            \draw[guide] (0,\gy) -- (43.5,\gy);
            \node[anchor=east,inner sep=1.7pt] at (0,\gy) {\tick};
        }
        \draw[axis] (0,19) -- (0,0) -- (43.5,0);
        \foreach \xx/\method/\rr/\mm/\aa in {2.55/FNO/42.27/39.47/66.03,
            7.45/U-Net/19.10/17.62/32.68,
            12.50/U-FNO/28.19/27.11/39.50,
            18.30/SAU-FNO/26.23/25.17/36.49,
            24.80/DeepOHeat/23.98/23.15/33.49,
            30.95/T/8.71/8.04/15.60,
            36.20/B/13.13/12.42/21.34,
            41.35/L/15.36/14.66/24.16}{
            \filldraw[bar,fill=rmsebar,draw=rmsebar!75!black] ({\xx-1.92},0) rectangle ++(1.12,{19*\rr/70});
            \filldraw[bar,fill=maebar,draw=maebar!75!black] ({\xx-0.56},0) rectangle ++(1.12,{19*\mm/70});
            \filldraw[bar,fill=maxaebar,draw=maxaebar!75!black] ({\xx+0.80},0) rectangle ++(1.12,{19*\aa/70});
            \node[model] at (\xx,-0.85) {\method};
        }
        \draw[axis] (28.8,-3.55) -- (43.3,-3.55);
        \node[model] at (36.05,-4.05) {Therm-FM};
    \end{scope}
    \begin{scope}[shift={(55,5.5)}]
        \node[charttitle] at (21.75,21.8) {(b) Few-shot adaptation ($K=10$)};
        \foreach \tick in {0,4,8,12}{
            \pgfmathsetmacro{\gy}{19*\tick/14}
            \draw[guide] (0,\gy) -- (43.5,\gy);
            \node[anchor=east,inner sep=1.7pt] at (0,\gy) {\tick};
        }
        \draw[axis] (0,19) -- (0,0) -- (43.5,0);
        \foreach \xx/\method/\rr/\mm/\aa in {2.55/FNO/3.30/2.67/10.91,
            7.45/U-Net/2.26/1.80/9.42,
            12.50/U-FNO/2.65/2.23/8.68,
            18.30/SAU-FNO/2.56/2.16/7.87,
            24.80/DeepOHeat/6.22/5.48/12.57,
            30.95/T/2.09/1.71/6.70,
            36.20/B/2.45/2.10/7.08,
            41.35/L/2.88/2.48/7.96}{
            \filldraw[bar,fill=rmsebar,draw=rmsebar!75!black] ({\xx-1.92},0) rectangle ++(1.12,{19*\rr/14});
            \filldraw[bar,fill=maebar,draw=maebar!75!black] ({\xx-0.56},0) rectangle ++(1.12,{19*\mm/14});
            \filldraw[bar,fill=maxaebar,draw=maxaebar!75!black] ({\xx+0.80},0) rectangle ++(1.12,{19*\aa/14});
            \node[model] at (\xx,-0.85) {\method};
        }
        \draw[axis] (28.8,-3.55) -- (43.3,-3.55);
        \node[model] at (36.05,-4.05) {Therm-FM};
    \end{scope}
    \end{tikzpicture}
    \caption{\scopeid{5} errors on the same 2,500-sample holdout:
(a) frozen zero-shot evaluation;
(b) adaptation with 10 labeled samples per package (50 total).
The panels use different vertical scales.}
\label{fig:s5-ood-combined}
\end{figure*}

\begin{table}[!t]
\caption{Per-package zero-shot RMSE (K) on the full \scopeid{5} test set. Each case contributes 1,000 samples.}
\label{tab:s5-per-case-main}
\centering
\footnotesize
\renewcommand{\arraystretch}{1.05}
\setlength{\tabcolsep}{3pt}
\begin{tabular*}{\columnwidth}{@{\extracolsep{\fill}}lrrrrr@{}}
\toprule
Method & Case11 & Case12 & Case13 & Case14 & Case15 \\
\midrule
FNO        & 94.58 & 20.02 & 50.96 & 20.19 & 25.12 \\
U-Net      & 16.17 & \textbf{3.25} & 40.36 & 17.44 & 17.05 \\
U-FNO      & 59.79 & 6.21 & 15.68 & 19.17 & 39.91 \\
SAU-FNO    & 55.46 & 4.71 & 37.18 & 9.88 & 23.38 \\
DeepOHeat  & 4.79 & 20.76 & 10.01 & 44.63 & 39.69 \\
Therm-FM-T & 8.42 & 9.94 & 11.17 & \textbf{7.50} & \textbf{6.33} \\
Therm-FM-B & \textbf{4.48} & 15.33 & \textbf{9.03} & 18.07 & 18.36 \\
Therm-FM-L & 6.15 & 13.05 & 12.76 & 18.80 & 25.92 \\
\bottomrule
\end{tabular*}
\end{table}

\subsection{Cross-Package OOD and Adaptation (\scopeid{5})}
\label{sec:level5-ood}

From \scopeid{4} to the full \scopeid{5} test set, the lowest MAE rises from 0.585~K for Therm-FM-L to 8.003~K for Therm-FM-T, a $13.7$-fold difference (Table~\ref{tab:progressive-results}; all six metrics in Table~\ref{tab:appendix-s5-all-metrics}). For Therm-FM-L alone, MAE rises from 0.585 to 14.633~K, approximately $25\times$ its \scopeid{4} value. Therm-FM-T leads all four reported \scopeid{5} metrics, reversing the \scopeid{4} ranking of the Therm-FM variants by RMSE and MAE.

The per-case RMSEs reveal package-dependent rankings (Table~\ref{tab:s5-per-case-main}). Therm-FM-T has the lowest aggregate RMSE, with per-case values of 6.33--11.17~K. It leads Cases~14 and~15, while Therm-FM-B leads Cases~11 and~13, and U-Net leads Case~12. DeepOHeat ranges from 4.79~K on Case~11 to 44.63~K on Case~14; U-Net ranges from 3.25~K on Case~12 to 40.36~K on Case~13.

\begin{figure}[!t]
    \centering
    \includegraphics[width=\columnwidth,height=40mm,keepaspectratio]{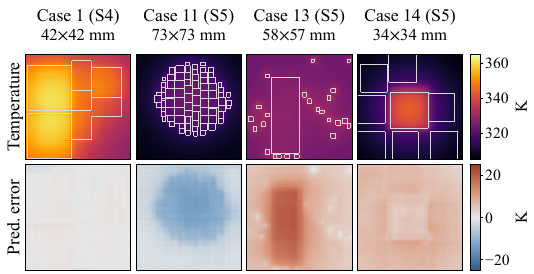}
    \caption{Temperature fields and prediction errors for
\scopeid{4} Case~1 (median-RMSE test sample) and
\scopeid{5} Cases~11, 13, and~14.
Top: HotSpot reference temperatures with chiplet outlines.
Bottom: signed errors ($\widehat{T}-T$) of frozen Therm-FM-T.}
    \label{fig:s5-samples}
\end{figure}

These cases combine geometry and power-distribution shifts. Case~11's $73\times73$~mm$^2$ interposer exceeds the source dimension limits of 59 and 61~mm along the two axes, while the nominal-power channel remains within its source range. The derived convection resistance also changes with heatsink area. \scopeid{5} therefore combines structural and physical input-distribution shifts. Figure~\ref{fig:s5-samples} shows broad underprediction by frozen Therm-FM-T in the Case~11 example and overprediction in the Case~13 and Case~14 examples. These signed errors can bias thermal-margin estimates, motivating inspection of spatial errors alongside aggregate metrics. Appendix~\ref{app:generated-viz} shows median-RMSE samples for all five held-out packages before and after adaptation.

With 10 labels per case, MAE falls by 76--93\% relative to each model's own frozen score on the common holdout (Table~\ref{tab:l5-adaptation}). FNO achieves the largest relative reduction, and Therm-FM-T achieves the lowest residual MAE of 1.71~K. Therm-FM-T has a 2.09~K RMSE and a 6.70~K mean worst-cell error at this budget (Fig.~\ref{fig:s5-ood-combined}(b)). At $K=50$, Therm-FM-T and Therm-FM-L are within 0.03~K of U-Net's best MAE of 1.01~K. At $K=500$, U-FNO achieves the lowest MAE of 0.59~K.

\subsection{Accuracy and Computational Cost Trade-offs}
\label{sec:accuracy-cost}

On \scopeid{4}, Therm-FM-L achieves 4.1\% lower RMSE than Therm-FM-B at $1.7\times$ the training cost (Tables~\ref{tab:baseline-configs} and~\ref{tab:progressive-results}). U-FNO comes within 16\% of Therm-FM-L's RMSE using less than 1\% of its parameters, with slightly higher training cost. U-Net achieves sub-kelvin MAE with 875 GPU$\cdot$s, making it a candidate for repeated training. The supervised DeepOHeat baseline trains fastest but has the highest within-family MAE. With one GPU, GPU$\cdot$s equals training wall time in seconds. Costs exclude pretraining, label generation, conversion, and hyperparameter exploration. These results show that a compact model can achieve competitive within-family accuracy, while parameter count alone does not determine training cost. Training costs for \scopeid{2} and \scopeid{3} are reported in Table~\ref{tab:appendix-training-cost}.

\section{Key Findings and Design Implications}
\label{sec:key-findings}

\textbf{First, low absolute errors remain attainable as the tasks
expand from fixed designs to within-family DSE.}
The \scopeid{1} results provide a low-error fixed-design
reference, while the best MAE increases from 0.248~K
in \scopeid{2} to 0.585~K in \scopeid{4}.
Even with joint variation in layout, materials, and cooling,
seven of the eight configurations retain MAE below 1~K
in \scopeid{4}.

\begin{table}[!t]
\caption{Target-domain adaptation on \scopeid{5}: MAE (K) on a fixed 2,500-sample holdout. $K$ counts labeled fields per unseen package ($5K$ total). $K=0$ is frozen zero-shot evaluation on this same holdout; Table~\ref{tab:progressive-results} uses all 5,000 \scopeid{5} samples. Complete curves for all six metrics and $K\in\{100,250\}$ are given in Appendix~\ref{app:ood-results} (Fig.~\ref{fig:appendix-fewshot-metrics}).}
\label{tab:l5-adaptation}
\centering
\footnotesize
\renewcommand{\arraystretch}{1.05}
\setlength{\tabcolsep}{4.8pt}
\begin{tabular*}{\linewidth}{@{\extracolsep{\fill}}lcccc@{}}
\toprule
Method & $K = 0$ & $K = 10$ & $K = 50$ & $K = 500$ \\
\midrule
U-FNO              & 27.11 & 2.23 & 1.06 & \textbf{0.59} \\
SAU-FNO            & 25.17 & 2.16 & 1.08 & 0.71 \\
FNO                & 39.47 & 2.67 & 1.17 & 0.72 \\
U-Net              & 17.62 & 1.80 & \textbf{1.01} & 0.73 \\
DeepOHeat          & 23.15 & 5.48 & 3.06 & 2.13 \\
Therm-FM-T (21M)    & \textbf{\phantom{0}8.04} & \textbf{1.71} & 1.04 & 0.72 \\
Therm-FM-B (158M)   & 12.42 & 2.10 & 1.24 & 0.62 \\
Therm-FM-L (629M)   & 14.66 & 2.48 & 1.02 & 0.67 \\
\bottomrule
\end{tabular*}
\end{table}

\textbf{Second, cross-package generalization presents a much
greater challenge.}
On the full \scopeid{5} test set, the best MAE reaches
8.003~K, $13.7\times$ the \scopeid{4} minimum.
This sharp increase on unseen packages contrasts with
the low absolute errors within known families,
identifying cross-package transfer as a key challenge
for current thermal surrogates.

\textbf{Third, limited target supervision can substantially
mitigate this gap.}
With 10 labeled fields per unseen package, adaptation
reduces MAE by 76--93\% relative to each model's frozen
baseline on the same holdout, with Therm-FM-T reaching
1.71~K.
These results support few-shot adaptation as an effective
option when a small budget of target simulations is available.

\section{Conclusion}
\label{sec:conclusion}

To the best of our knowledge, OpenTherm is the first
open dataset and benchmark for 2.5D/3D-IC thermal prediction
that jointly supports fixed-design prediction,
within-family design exploration, and package-disjoint transfer.
Our evaluation shows that most models retain low absolute
errors under broader within-family variation, but degrade
sharply on unseen packages.
Few-shot adaptation substantially reduces this transfer gap,
highlighting the value of limited target supervision.
By providing shared datasets, standardized evaluation protocols,
and extensible interfaces, OpenTherm establishes a reproducible
foundation for model training and fair comparison in
learning-based thermal prediction.

\bibliographystyle{IEEEtran}
\bibliography{references}

\appendices

\section{Benchmark Reporting and Fair-Comparison Contract}
\label{app:fair-comparison}

\benchname{} results are directly comparable on the primary benchmark
only when they use the same dataset version, immutable splits, input and
label definitions, preprocessing, normalization, and reference metrics.
Training hardware, parameter count, optimization configuration, and
result provenance must also be reported.
Frozen zero-shot evaluation and target-domain adaptation are separate
tracks, and any alternative simulator, resolution, split, preprocessing,
or metric must be reported as a separate track.
Table~\ref{tab:reporting} summarizes the minimum reporting record.

\input{appendix_new}

\end{document}

%% file: appendix_new.tex
\begin{table}[!t]
\caption{Minimum record for a comparable benchmark result.}
\label{tab:reporting}
\centering
\small
\setlength{\tabcolsep}{2.8pt}
\renewcommand{\arraystretch}{1.06}
\begin{tabular}{@{}p{0.22\columnwidth}p{0.70\columnwidth}@{}}
\toprule
Item & Required record \\
\midrule
Data & Version, case list, split hash \\
Labels & Simulator, version, grid, tolerances, boundary model \\
Inputs & Channel semantics, normalization, resolution handling \\
Zero-shot & Frozen checkpoint and unchanged preprocessing \\
Adaptation & $K$ per case, pool/holdout split, nested indices \\
Metrics & RMSE, MAE, $R^2$, MaxAE, $E_{T_{\max}}$, \topmae \\
Optimization & Loss, learning rate, epochs, checkpoint rule \\
Compute & Device, GPU count, wall time, GPU$\cdot$s, parameters \\
Provenance & Author-reported, released-code, or reimplementation \\
\bottomrule
\end{tabular}
\end{table}

\section{Physical Formulation and Dataset Specification}
\label{app:data-specification}

\subsection{Thermal Formulation}
\label{app:thermal-formulation}

Let $\Omega$ denote the package domain, $T(\mathbf{x},t)$ its temperature,
$\rho$ the mass density, $C_p$ the specific heat, $k$ the thermal
conductivity, and $Q$ the volumetric heat source. The transient field obeys
\begin{equation}
\rho C_p\frac{\partial T}{\partial t}
-\nabla\!\cdot\!\left(k\nabla T\right)=Q
\quad \text{in }\Omega.
\label{eq:app-heat}
\end{equation}
Steady-state analysis removes the temporal term. On the boundary, the
reference problem may impose
\begin{align}
T &= T_D && \text{on }\Gamma_D, \\
-\mathbf{n}\!\cdot k\nabla T &= q_N && \text{on }\Gamma_N, \\
-\mathbf{n}\!\cdot k\nabla T &= h(T-T_{\mathrm{amb}})
&& \text{on }\Gamma_h,
\end{align}
for Dirichlet, prescribed-flux, and convective conditions. Across a material
interface $\Gamma_I$, temperature and normal heat flux remain continuous,
\begin{equation}
T^- = T^+,
\qquad
\mathbf{n}\!\cdot k^-\nabla T^-
=\mathbf{n}\!\cdot k^+\nabla T^+.
\end{equation}
Transient tasks additionally specify $T(\mathbf{x},0)=T_0(\mathbf{x})$.
For a fixed structural configuration $\mathcal{M}$, these equations define a
solution operator from power, material, boundary, and temporal variables to
the temperature field. \scopeid{1} preserves both steady and transient source
tasks; generated \scoperange{2}{5} tasks are steady state.

\subsection{HotSpot Generation Configuration}
\label{app:hotspot}

All generated \scoperange{2}{5} labels use the open-source HotSpot
grid-based detailed-3D flow at $64\times64$ resolution. The exact executable,
configuration files, layer files, units, and conversion scripts are released
with the benchmark artifact. Temperatures are exported in kelvin. The six
layer stack is fixed across \scoperange{2}{5}; only the benchmark variables
listed below change.

\begin{table*}[!t]
\caption{Six-layer HotSpot stack used for generated \scoperange{2}{5} tasks.}
\label{tab:appendix-stack}
\centering
\setlength{\tabcolsep}{5.5pt}
\renewcommand{\arraystretch}{1.08}
\begin{tabular}{@{}clcrrrr@{}}
\toprule
\# & Layer (bottom to top) & Power & Heat capacity & Resistivity & Conductivity & Thickness \\
\midrule
0 & Substrate
  & No & $1.06\times10^6$ & 3.33 & $\approx0.30$ & 200 $\mu$m \\

1 & Epoxy SiO$_2$ underfill + C4
  & No & $2.32\times10^6$ & 0.625 & $\approx1.6$ & 70 $\mu$m \\

2 & Silicon interposer
  & No & $1.75\times10^6$ & 0.01 & 100 & 110 $\mu$m \\

3 & Underfill with microbumps
  & No & $2.32\times10^6$ & 0.625 & $\approx1.6$ & 10 $\mu$m \\

4 & Chip silicon / filler
  & Yes & $1.75\times10^6$ & 0.01 / variable & 100 / 0.2--5.0 & 150 $\mu$m \\

5 & TIM
  & No & $4.00\times10^6$ & 0.25 & 4.0 & 20 $\mu$m \\
\bottomrule
\end{tabular}
\end{table*}

\subsection{Parameter Distributions}
\label{app:sampling-ranges}

Sampling distributions are specified in Section~\ref{sec:data-generation};
Table~\ref{tab:appendix-ranges} lists the resulting empirical ranges of the
model inputs.

\begin{table*}[!t]
\caption{Empirical model-input ranges.
Dashes denote absent channels.
For the cell-wise conductivity channel $k$, the upper value
$100$~W/(m$\cdot$K) comes from fixed silicon cells; only filler
conductivity is sampled, over $[0.2,5.0]$~W/(m$\cdot$K).}
\label{tab:appendix-ranges}
\centering
\setlength{\tabcolsep}{8pt}
\renewcommand{\arraystretch}{1.08}
\begin{tabular}{@{}lrrrr@{}}
\toprule
Channel & \scopeid{2} & \scopeid{3} & \scopeid{4} & \scopeid{5} OOD \\
\midrule
Nominal chiplet power (W) & 0--300 & 0--300 & 0--300 & 0--280 \\
Physical $x$ (mm) & 0--59 & 0--59 & 0--59 & 0--72.7 \\
Physical $y$ (mm) & 0--61 & 0--61 & 0--61 & 0--72.7 \\
$k$ (W/(m$\cdot$K)) & -- & 0.2--100 & 0.2--100 & 0.2--100 \\
$T_{\mathrm{amb}}$ (K) & -- & -- & 298.2--328.1 & 298.2--328.1 \\
$h$ (W/(m$^2$K)) & -- & -- & 500--19,990 & 500--20,000 \\
$R_{\mathrm{conv}}$ (K/W) & -- & -- & 0.00087--0.2081 & 0.00059--0.1080 \\
\bottomrule
\end{tabular}
\end{table*}

\subsection{Input Tensor Semantics}
\label{app:model-channels}

The cumulative model inputs follow Eq.~\eqref{eq:channels}, and
Table~\ref{tab:model-input-channels} lists their released names, units, and
semantics. Conductivity is a cell-wise material field, while
$T_{\mathrm{amb}}$, $h$, and $R_{\mathrm{conv}}$ are sample-global scalars
broadcast over the grid. Each input channel is normalized independently using
statistics fitted on the training subset only.

\begin{table*}[!t]
\caption{Model-input channels for generated \scoperange{2}{5} tasks.}
\label{tab:model-input-channels}
\centering
\setlength{\tabcolsep}{4.2pt}
\renewcommand{\arraystretch}{1.08}
\begin{tabular}{@{}cllccc p{5.7cm}@{}}
\toprule
\# & Channel & Unit & $P=3$ & $P=4$ & $P=7$ & Semantics \\
\midrule
0 & \texttt{chiplet\_power} & W
  & Yes & Yes & Yes
  & Nominal chiplet power broadcast over occupied cells; zero in background \\

1 & \texttt{grid\_x} & mm
  & Yes & Yes & Yes
  & Physical $x$ coordinate \\

2 & \texttt{grid\_y} & mm
  & Yes & Yes & Yes
  & Physical $y$ coordinate \\

3 & \texttt{local\_thermal\_k} & W/(m$\cdot$K)
  & -- & Yes & Yes
  & Cell-wise material conductivity \\

4 & \texttt{ambient\_K} & K
  & -- & -- & Yes
  & Sample-global ambient temperature \\

5 & \texttt{h\_w\_m2k} & W/(m$^2$K)
  & -- & -- & Yes
  & Sample-global convection coefficient \\

6 & \texttt{r\_convec\_k\_per\_w} & K/W
  & -- & -- & Yes
  & Derived convective resistance \\
\bottomrule
\end{tabular}
\end{table*}

Converted tensors use \texttt{float32}, a $64\times64$ grid, and a retained
singleton depth $Z=1$. The released CSVs additionally retain
chiplet identifiers, occupancy and edge masks, and normalized coordinates for
reconstruction, but these auxiliary columns are not given to the evaluated
models.

\subsection{Layout Generation}
\label{app:layout-generation}

Layouts are sampled in the four modes and ratio given in
Section~\ref{sec:data-generation}, varying legal position and orientation.
Accepted chiplets are rasterized onto the $64\times64$ grid; occupied cells
inherit chiplet identity and nominal power, while background cells remain
zero. The released case manifest and generator configuration define the exact
legal-layout constraints.

\section{Splits, OOD Protocol, and Metrics}
\label{app:evaluation-contract}

\subsection{Case-Balanced \scoperange{2}{4} Splits}

Split sizes are given in Section~\ref{sec:evaluation-protocol}. Samples are
interleaved round-robin by case, and normalization statistics are fitted only
on the 10,800-sample training subset and remain fixed for validation and test.

\subsection{Official \scopeid{5} Zero-Shot Protocol}

Cases~11--15 are excluded from source training, validation, normalization
fitting, and checkpoint selection. The official zero-shot track evaluates the
frozen \scopeid{4} checkpoint on all 5,000 \scopeid{5} samples with unchanged
preprocessing and no target labels. Because an unseen package can shift both
structure and the empirical support of model-visible coordinates or derived
variables, this track is termed \emph{Cross-Package OOD}, not a controlled
structural-only intervention.

\begin{table*}[!t]
\caption{Principal stress characteristics of held-out \scopeid{5} systems.
Each case primarily stresses the listed attribute but is not a strict
single-factor intervention.}
\label{tab:l5-ood}
\centering
\setlength{\tabcolsep}{4.0pt}
\begin{tabular}{clp{10.2cm}}
\toprule
Case & Principal stress & Difference from Cases~1--10 \\
\midrule
11 & Chiplet count & 80 chiplets versus a training maximum of 61; utilization is reduced to 32\% for feasible generation \\
12 & Power density & 0.85 W/mm$^2$ versus a training maximum of 0.745 W/mm$^2$ \\
13 & Size heterogeneity & One $42\times16$ mm die plus 24 micro-chiplets of 2--2.6 mm \\
14 & Power concentration & One chiplet contributes 90.9\% of total power versus a training maximum of 69.2\% \\
15 & Package utilization & 68.1\% utilization versus a training maximum of 67.1\% \\
\bottomrule
\end{tabular}
\end{table*}

\subsection{Target-Domain Adaptation Protocol}
\label{app:adaptation-protocol}

For each held-out case, the first 500 released samples form an adaptation
pool and the final 500 form a fixed evaluation holdout. The released nested
index lists define $K\in\{10,50,100,250,500\}$ labels per case, so the total
target-label budget is $5K$. Each run starts from the frozen \scopeid{4}
checkpoint, reinitializes the optimizer, retains the source normalization,
and fine-tunes for 50 epochs. Non-Therm-FM models use Adam with learning rate
$10^{-4}$, weight decay $10^{-4}$, and batch size 20; Therm-FM uses
$1.5\times10^{-5}$, one tenth of its training rate, because the larger rate
used by the operator baselines is unstable for its pretrained stack. The
target holdout is never used to fit normalization or update weights. The reported tables use
the epoch-50 checkpoint.

The adaptation curve's $K=0$ value is the frozen checkpoint evaluated on the
common 2,500-sample holdout. It therefore differs slightly from the official
zero-shot score, which uses all 5,000 \scopeid{5} samples; keeping the two
denominators separate prevents a test-set change from appearing as an
adaptation gain.

\subsection{Metric Definitions}
\label{app:metric-definitions}

Per-sample metrics are defined in Eq.~\eqref{eq:error-metrics} and averaged
over the test set as described in Section~\ref{sec:evaluation-protocol}.
All eight baselines share one metric implementation. For a single output
channel, as in \scoperange{2}{5}, the pooled coefficient of determination is
\begin{equation}
R^2=1-\frac{\sum_{j,i}(\hat{T}_{j,i}-T_{j,i})^2}
{\sum_{j,i}(T_{j,i}-\bar{T})^2},
\end{equation}
where $\bar{T}$ is the pooled reference mean; multi-channel outputs average
this score over channels.

\section{Baseline and Training Details}
\label{app:training-details}

\subsection{Baseline Architectures}

\begin{table*}[!t]
\caption{Model configurations and input-dependent parameter counts for
\scoperange{2}{4} ($P=3/4/7$).}
\label{tab:appendix-architecture}
\centering
\small
\setlength{\tabcolsep}{4.5pt}
\renewcommand{\arraystretch}{1.08}
\begin{tabular}{@{}lp{0.43\textwidth}rrr@{}}
\toprule
Model & Configuration & $P=3$ & $P=4$ & $P=7$ \\
\midrule
U-FNO
& \texttt{Net3d}; modes $(10,10,1)$; width 36
& 5,103,905 & 5,103,941 & 5,104,049 \\

SAU-FNO
& \texttt{SAUNet3d}; modes $(10,10,1)$; width 36
& 5,107,901 & 5,107,937 & 5,108,045 \\

FNO
& \texttt{FNO3d}; modes $(12,12,1)$; width 72
& 11,974,721 & 11,974,793 & 11,975,009 \\

U-Net
& Channels $(58,116,232,348)$; three max-pools; nearest-neighbor upsampling;
skip connections; no BN/dropout
& 5,060,037 & 5,060,559 & 5,062,125 \\

DeepOHeat
& Frozen 128-D Gaussian Fourier coordinate trunk; 128-wide 3-layer SiLU trunk;
flattened-input 256-wide 7-layer branch; inner product + bias
& 3,721,985 & 4,770,561 & 7,916,289 \\

Therm-FM-T/B/L
& scOT/Poseidon pretrained checkpoints
& \multicolumn{3}{c}{20.7M / 157.6M / 628.4M} \\
\bottomrule
\end{tabular}
\end{table*}

Because $Z=1$, the third Fourier mode count is one. Increasing the channel
count changes only the input layer for convolutional and Fourier baselines,
but substantially enlarges DeepOHeat's flattened branch.

\subsection{Optimization Configuration}

\begin{table*}[!t]
\caption{Training configuration. All normalizers are fitted on training data
only and applied per channel.}
\label{tab:appendix-optimization}
\centering
\setlength{\tabcolsep}{4pt}
\renewcommand{\arraystretch}{1.08}
\begin{tabular}{@{}p{0.14\textwidth}p{0.30\textwidth}p{0.30\textwidth}@{}}
\toprule
Setting & FNO/U-FNO/SAU-FNO/U-Net/DeepOHeat & Therm-FM-T/B/L \\
\midrule
Optimizer
& Adam
& AdamW \\

Learning Rate
& $10^{-3}$
& $1.5\times10^{-4}$ (embedding/recovery: $1.5\times10^{-3}$) \\

Weight decay
& $10^{-4}$
& $10^{-6}$ \\

Schedule
& StepLR(step 2, $\gamma=0.9$)
& Cosine; zero warmup \\

Batch size
& 20
& 40 \\

Epochs
& 100
& 100 \\

Objective
& Normalized-space MSE
& Normalized-space MSE divided by mean squared target; max grad norm 5 \\

Checkpoint
& Final epoch
& Validation-best \\

Hardware
& One RTX A6000 48GB
& One RTX A6000 48GB \\
\bottomrule
\end{tabular}
\end{table*}

All evaluated methods are label-supervised. For DeepOHeat this differs from
the original residual-only training formulation and keeps the benchmark
question fixed: given identical simulated labels, which model best predicts
the temperature field? FNO, U-FNO, SAU-FNO, U-Net, and DeepOHeat use the
final-epoch checkpoint with early stopping disabled; Therm-FM retains the
validation-best checkpoint selected by its scOT training pipeline.

\subsection{Software and Hardware}

Experiments use NVIDIA RTX A6000 48GB GPUs and an AMD EPYC 7543 CPU with
PyTorch~2.0.1 and CUDA~11.8. All baselines use a single GPU. Wall time covers model training
only and excludes one-time HotSpot generation, format conversion, checkpoint
download, and manual hyperparameter exploration.

\subsection{Executable Interface and Model Extension}
\label{app:interface}

The public benchmark repository (\url{https://github.com/Day333/OpenTherm}) exposes one entry point for
training, Within-Family DSE testing, frozen Cross-Package OOD evaluation, and
target-case adaptation.

Every method is adapted to the same repository-level tensor contract: each
sample pairs a $P$-channel input field with a temperature field on the same
$64\times64\times Z$ grid, where the current generated release uses $Z=1$ and
$P\in\{3,4,7\}$.  A conventional PyTorch model needs only an implementation
and a registry entry that declares its recipe and channel-dependent
constructor; the shared runner then supplies the immutable split,
normalization, six metrics, JSON output, and \scopeid{5} adaptation loop.
Adding a conventional PyTorch baseline requires one implementation file and
one short \texttt{MODEL\_ZOO} entry containing its metric prefix, optimizer
recipe, adaptation rate, and a constructor that receives $P$, $Z$, and grid
size.  Models with custom training stacks use the same tensor and metric
contracts through a separate adapter.

This interface separates architecture-specific tensor conversion and
optimization from benchmark-controlled data and evaluation.  A released
result can therefore be regenerated with the same sample identifiers and
metric implementation, while a new model can be compared without copying or
silently modifying the evaluation code.

\section{\scopeid{1} Source-Suite Results}
\label{app:level1}

\begin{figure*}[!tbp]
  \centering
  \includegraphics[width=0.66\textwidth]{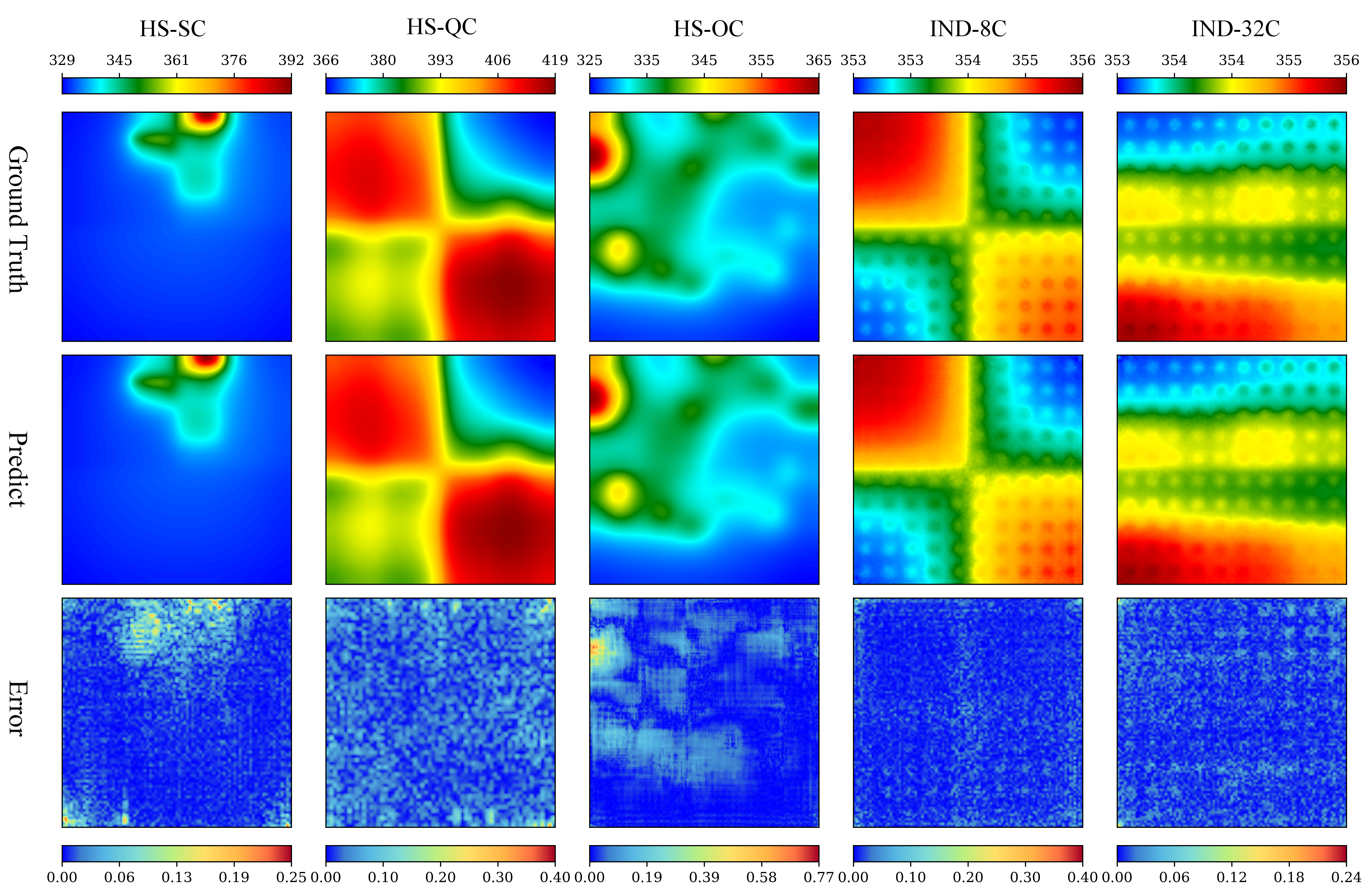}\\[-2pt]
  {\footnotesize\textbf{(a)} Steady-state HS-SC/HS-QC/HS-OC and IND-8C/IND-32C results.}\par\vspace{3pt}
  \includegraphics[width=0.66\textwidth]{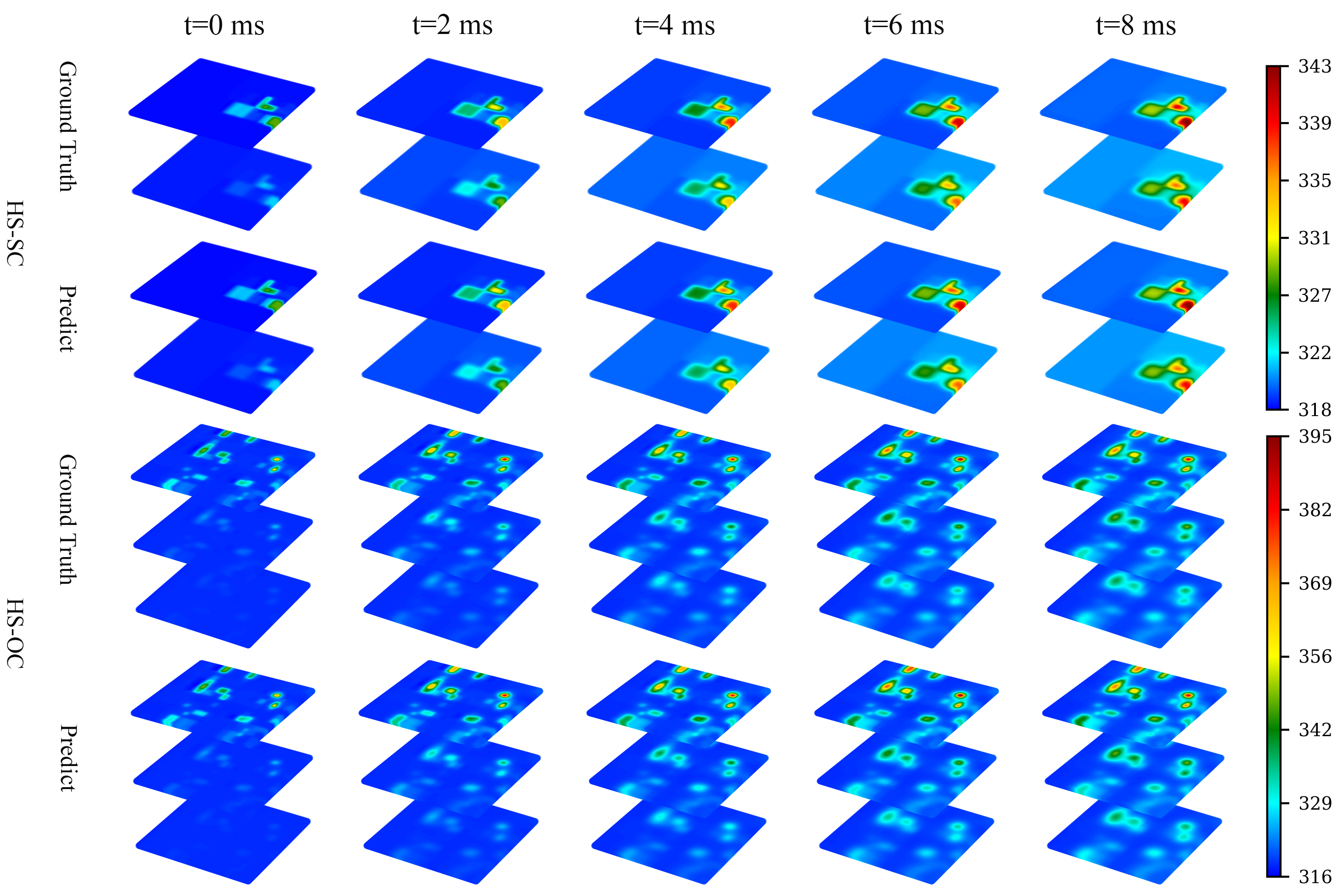}\\[-2pt]
  {\footnotesize\textbf{(b)} Transient HS-SC and HS-OC results across sampled time steps.}
  \caption{Qualitative \scopeid{1} visualizations reproduced directly from
  the best-performing Therm-FM source configurations~\cite{huang2026thermfm};
  they were not regenerated for this paper.}
  \label{fig:appendix-l1-visualizations}
\end{figure*}

\scopeid{1} comprises eleven fixed-design source tasks, including
steady-state and transient Alpha EV6 configurations and two industrial
packages. Table~\ref{tab:s1-summary} summarizes the S1 results: the Therm-FM
values coincide with those reported in the Therm-FM
study~\cite{huang2026thermfm}, whereas the other baselines are trained and
evaluated in the current benchmark. Per-task tables are omitted for
conciseness. Figure~\ref{fig:appendix-l1-visualizations} follows the Therm-FM naming:
HS-SC, HS-QC, and HS-OC denote the three HotSpot-simulated Alpha EV6
configurations (the EV6 tasks in Table~\ref{tab:s1-summary}), and IND-8C and
IND-32C denote the two industrial packages.

\paragraph{Qualitative observations.}
Figure~\ref{fig:appendix-l1-visualizations}(a) shows that the selected
Therm-FM results preserve the global temperature distribution and dominant
hotspot locations across the three HotSpot cases and two industrial packages;
the visible residuals remain localized mainly near sharper thermal transitions.
Panel~(b) shows close agreement through the sampled transient steps for HS-SC
and HS-OC, including the temperature rise and spatial diffusion from 0 to
8~ms. These observations are qualitative and refer only to the source Therm-FM
visualizations; the corresponding quantitative results are reported in the
Therm-FM study cited above.

\section{Complete Within-Family DSE Results (\scoperange{2}{4})}
\label{app:insupport-results}

The main paper retains four metrics per Scope to expose the central trend:
RMSE and MAE for global field accuracy, together with worst-cell MaxAE and
Top-50 MAE.
Tables~\ref{tab:appendix-s2-full}--\ref{tab:appendix-s4-full} restore all six
reported metrics for every baseline.  The datasets are independently
generated under a common case-balanced protocol; the progression should
therefore be read as increasing represented-distribution difficulty, not as a
paired causal ablation on the same samples.

\begin{table*}[!t]
\caption{Complete \scopeid{2} Within-Family DSE results under represented layout
variation.  Errors are in kelvin;
higher $R^2$ is better and lower is better for all other metrics.}
\label{tab:appendix-s2-full}
\centering
\setlength{\tabcolsep}{8.5pt}
\begin{tabular}{lrrrrrr}
\toprule
Method & RMSE & MAE & $R^2$ & MaxAE & $E_{T_{\max}}$ & Top-50 MAE \\
\midrule
FNO         & 0.8284 & 0.6207 & 0.9977 & 3.9652 & 0.4667 & 0.4943 \\
U-Net       & 0.7375 & 0.5412 & 0.9983 & 5.2731 & 0.4492 & 0.4996 \\
U-FNO       & 0.4442 & 0.3265 & 0.9994 & 2.7572 & 0.2975 & 0.2978 \\
SAU-FNO     & 0.5336 & 0.4086 & 0.9991 & 2.6318 & 0.3779 & 0.4282 \\
DeepOHeat   & 3.7055 & 2.9376 & 0.9593 & 10.2608 & 2.8308 & 3.6447 \\
Therm-FM-T  & 0.5752 & 0.4102 & 0.9986 & 3.1207 & 0.3691 & 0.3992 \\
Therm-FM-B  & 0.4610 & 0.3198 & 0.9990 & 2.7091 & 0.2870 & 0.3114 \\
Therm-FM-L  & \textbf{0.3536} & \textbf{0.2477} & \textbf{0.9995} & \textbf{2.1448} & \textbf{0.2164} & \textbf{0.2378} \\
\bottomrule
\end{tabular}
\end{table*}

\paragraph{Reading \scopeid{2}.}
Therm-FM-L leads all six metrics, with U-FNO and Therm-FM-B close behind
on the global field errors; among the operator baselines U-FNO leads SAU-FNO
on every metric except MaxAE, and model scale helps Therm-FM monotonically
here.  DeepOHeat trails by a wide margin.  U-Net ranks lower on the
worst-cell and hotspot-region errors (MaxAE and Top-50 MAE) than on global RMSE,
which also illustrates why hotspot-region and field-wide errors should be
reported together.

\begin{table*}[!t]
\caption{Complete \scopeid{3} Material Sweep results under Within-Family
DSE.}
\label{tab:appendix-s3-full}
\centering
\setlength{\tabcolsep}{8.5pt}
\begin{tabular}{lrrrrrr}
\toprule
Method & RMSE & MAE & $R^2$ & MaxAE & $E_{T_{\max}}$ & Top-50 MAE \\
\midrule
FNO         & 0.9619 & 0.7200 & 0.9967 & 4.4718 & 0.4826 & 0.5059 \\
U-Net       & 0.8432 & 0.6302 & 0.9975 & 5.0203 & 0.4753 & 0.5017 \\
U-FNO       & \textbf{0.6700} & \textbf{0.4928} & \textbf{0.9984} & \textbf{3.3669} & \textbf{0.3312} & \textbf{0.3730} \\
SAU-FNO     & 0.7078 & 0.5255 & 0.9982 & 3.6997 & 0.4080 & 0.4466 \\
DeepOHeat   & 3.9699 & 3.2259 & 0.9437 & 10.5532 & 3.5631 & 4.8260 \\
Therm-FM-T  & 0.9105 & 0.6787 & 0.9968 & 4.2679 & 0.4989 & 0.5507 \\
Therm-FM-B  & 0.7215 & 0.5165 & 0.9980 & 3.7171 & 0.3378 & 0.3781 \\
Therm-FM-L  & 0.7366 & 0.5268 & 0.9979 & 3.7848 & 0.3436 & 0.3904 \\
\bottomrule
\end{tabular}
\end{table*}

\paragraph{Reading \scopeid{3}: Material Sweep.}
U-FNO leads all six metrics once conductivity becomes an explicit input
field, ahead of SAU-FNO and every Therm-FM variant.  The best RMSE rises by
about 89\% relative to \scopeid{2}, yet the best $R^2$ remains above 0.998, so
represented material variation increases difficulty without causing a
generalization break.  Model scale is still not
a reliable predictor: Therm-FM-L does not improve over Therm-FM-B here.

\paragraph{Metric-wise behavior across represented support.}
The six metrics expose a stable distinction between reconstructing the full
field and tracking the thermally critical region.  SAU-FNO is never the
best global predictor, yet its MaxAE is the second best on every
within-family Scope; a deployment concerned mainly with bounding the
worst-case local error could therefore reach a different model choice from
one that requires an accurate temperature field everywhere.  Conversely, DeepOHeat's
global and hotspot errors deteriorate together as input dimensionality grows,
suggesting that its present flattened branch is a limiting factor under the
$64\times64$ representation rather than a metric-specific failure.  U-FNO
and SAU-FNO remain comparatively balanced across all six criteria.

\paragraph{Implication for progressive evaluation.}
No architecture should be described as generally superior from one Scope.
The leading model changes with the Scope (Therm-FM-L at \scopeid{2} and
\scopeid{4}, U-FNO at \scopeid{3}), the best Therm-FM variant switches from L
to B at \scopeid{3}, and --- as \scopeid{5} shows next --- the
best within-family model is not the best cross-package one.  This is
informative: the configuration that best absorbs represented placement,
material, and cooling variation need not be the one that transfers to
unseen packages.  The progression therefore functions as a capability
profile rather than a single scalar leaderboard.  The first Scope at which
error rises sharply indicates where performance begins to degrade as broader
physical variation is introduced.

\begin{table*}[!t]
\caption{Complete \scopeid{4} Cooling Co-Design results under Within-Family
DSE.}
\label{tab:appendix-s4-full}
\centering
\setlength{\tabcolsep}{8.5pt}
\begin{tabular}{lrrrrrr}
\toprule
Method & RMSE & MAE & $R^2$ & MaxAE & $E_{T_{\max}}$ & Top-50 MAE \\
\midrule
FNO         & 1.2912 & 0.9939 & 0.9977 & 5.4362 & 0.6361 & 0.6929 \\
U-Net       & 1.2713 & 0.9764 & 0.9978 & 7.0174 & 0.8183 & 0.8383 \\
U-FNO       & 0.9391 & 0.7237 & 0.9988 & 5.0106 & 0.5180 & 0.5727 \\
SAU-FNO     & 0.9532 & 0.7394 & 0.9988 & 4.1366 & 0.6504 & 0.6826 \\
DeepOHeat   & 4.7394 & 3.9033 & 0.9699 & 11.7546 & 4.7950 & 5.9236 \\
Therm-FM-T  & 1.0496 & 0.7940 & 0.9985 & 4.8432 & 0.5989 & 0.6426 \\
Therm-FM-B  & 0.8462 & 0.6180 & 0.9990 & 4.2598 & 0.4095 & 0.4614 \\
Therm-FM-L  & \textbf{0.8113} & \textbf{0.5852} & \textbf{0.9991} & \textbf{4.0345} & \textbf{0.3696} & \textbf{0.4314} \\
\bottomrule
\end{tabular}
\end{table*}

\paragraph{Reading \scopeid{4}: Cooling Co-Design.}
Therm-FM-L leads all six metrics, with U-FNO the strongest operator
baseline on the global-field errors and SAU-FNO on MaxAE.  Relative
to \scopeid{2}, the best RMSE increases by roughly $2.3\times$, while the best
peak-temperature error stays below 0.5~K.  The distinction matters for
deployment: a predictor may be sufficiently accurate for thermal-limit
screening even when its full-field error has increased.  Most importantly,
this moderate within-family degradation is qualitatively smaller than the
case-disjoint failure reported next.

\section{Complete Cross-Package OOD and Adaptation Results}
\label{app:ood-results}

This section reports the complete \scopeid{5} results behind the condensed
main-paper discussion, under the official zero-shot protocol of
Appendix~\ref{app:evaluation-contract}.

The aggregate and per-case views answer complementary questions.  The
aggregate table measures expected performance over the released stress set
and supports a compact leaderboard.  The per-case RMSEs (Table~\ref{tab:s5-per-case-main}) reveal whether that
average reflects consistent transfer or success on one system offsetting
failure on another.  We report both because the five held-out systems stress
different combinations of chiplet count, die outline, size mixture, power
density, power concentration, and utilization; they are not repeated samples
from one homogeneous OOD population.

\begin{table*}[!t]
\caption{Complete zero-shot results on the official \scopeid{5} test set.
All error metrics are in kelvin; higher $R^2$ is better and lower is better
for the other metrics.}
\label{tab:appendix-s5-all-metrics}
\centering
\setlength{\tabcolsep}{7.2pt}
\begin{tabular}{lrrrrrr}
\toprule
Method & RMSE & MAE & $R^2$ & MaxAE & $E_{T_{\max}}$ & Top-50 MAE \\
\midrule
FNO         & 42.1755 & 39.3575 & -1.6429 & 65.9182 & 59.9215 & 39.7131 \\
U-Net       & 18.8542 & 17.3739 & 0.3776 & 32.4320 & 24.0189 & 18.0644 \\
U-FNO       & 28.1538 & 27.0604 & -0.2399 & 39.4894 & 31.6274 & 26.6474 \\
SAU-FNO     & 26.1208 & 25.0555 & -0.1239 & 36.3580 & 31.7934 & 26.9022 \\
DeepOHeat   & 23.9735 & 23.1409 & 0.0759 & 33.4330 & 25.6216 & 24.0784 \\
Therm-FM-T  & \textbf{8.6729} & \textbf{8.0029} & \textbf{0.9101} & \textbf{15.5347} & \textbf{10.4522} & \textbf{9.0454} \\
Therm-FM-B  & 13.0554 & 12.3390 & 0.6921 & 21.2864 & 16.7420 & 15.7459 \\
Therm-FM-L  & 15.3361 & 14.6325 & 0.6835 & 24.0876 & 17.1175 & 13.7678 \\
\bottomrule
\end{tabular}
\end{table*}

No single baseline dominates every held-out package.  Therm-FM-T wins
the pooled six-metric comparison and is the best model on C14 and C15, but
its advantage comes mainly from comparatively balanced errors
(6.3--11.2~K on every case) rather than from winning every case: Therm-FM-B
is best on C11 and C13 and U-Net on C12, showing that the aggregate OOD gap
combines several distinct package-level shifts. DeepOHeat
is also strong on the high-chiplet-count C11 case (4.79~K) but
degrades on C12, C14, and C15; SAU-FNO and U-Net show the opposite kind
of specialization, with their lowest errors on the high-power-density C12
case.  Moreover, larger Therm-FM variants transfer progressively worse
on five of the six metrics (all but Top-50 MAE): zero-shot RMSE rises from 8.67~K (T) to 13.06~K (B) and 15.34~K (L).  These ranking reversals show that strong within-family accuracy under
represented physical variation does not imply robust cross-package
generalization.

\subsection{Target-Case Adaptation}
\label{app:adaptation}

Adaptation follows the protocol in Appendix~\ref{app:evaluation-contract}; as
noted there, the $K=0$ column uses the common 2,500-sample holdout rather than
the official 5,000-sample zero-shot set of
Table~\ref{tab:appendix-s5-all-metrics}.

\paragraph{Reading the label budget.}
Because $K$ is counted per held-out case, even the smallest budget observes
every target case.  This protocol tests low-cost target supervision,
not zero-shot generalization, and the two results must not be conflated.  A method that improves the frozen S5 score addresses cross-package
transfer, whereas one that reaches a lower adapted score with the same
$K$ addresses target-domain sample efficiency. The complete curve is required because those
rankings need not agree.

\begin{table*}[!t]
\caption{RMSE (K) after target-case adaptation.  Each $K$ is the number of
labeled adaptation samples per held-out case.}
\label{tab:appendix-adaptation-rmse}
\centering
\setlength{\tabcolsep}{8.2pt}
\begin{tabular}{lrrrrrr}
\toprule
Method & $K=0$ & $K=10$ & $K=50$ & $K=100$ & $K=250$ & $K=500$ \\
\midrule
U-FNO       & 28.19 & 2.65 & 1.29 & \textbf{1.06} & 1.10 & \textbf{0.76} \\
SAU-FNO     & 26.23 & 2.56 & 1.32 & 1.20 & \textbf{0.86} & 0.90 \\
FNO         & 42.27 & 3.30 & 1.47 & 1.24 & 1.01 & 0.92 \\
U-Net       & 19.10 & 2.26 & 1.29 & 1.21 & 0.95 & 0.93 \\
DeepOHeat   & 23.98 & 6.22 & 3.69 & 3.28 & 2.91 & 2.71 \\
Therm-FM-T  & \textbf{8.71} & \textbf{2.09} & 1.32 & 1.10 & 0.92 & 0.92 \\
Therm-FM-B  & 13.13 & 2.45 & 1.48 & 1.18 & 0.99 & 0.81 \\
Therm-FM-L  & 15.36 & 2.88 & \textbf{1.27} & \textbf{1.06} & 0.92 & 0.86 \\
\bottomrule
\end{tabular}
\end{table*}

\begin{table*}[!t]
\caption{Complete target-case adaptation results for (a) MAE (K) and
(b) pooled $R^2$.  Each $K$ is the number of labels per held-out case.}
\label{tab:appendix-adaptation-mae-r2}
\centering
\scriptsize
\begin{minipage}[t]{0.49\textwidth}
\centering
\textbf{(a) MAE (K); lower is better.}\par\vspace{2pt}
\resizebox{\linewidth}{!}{%
\begin{tabular}{lrrrrrr}
\toprule
Method & 0 & 10 & 50 & 100 & 250 & 500 \\
\midrule
U-FNO       & 27.11 & 2.23 & 1.06 & 0.86 & 0.92 & \textbf{0.59} \\
SAU-FNO     & 25.17 & 2.16 & 1.08 & 0.98 & \textbf{0.69} & 0.71 \\
FNO         & 39.47 & 2.67 & 1.17 & 0.98 & 0.80 & 0.72 \\
U-Net       & 17.62 & 1.80 & \textbf{1.01} & 0.97 & 0.75 & 0.73 \\
DeepOHeat   & 23.15 & 5.48 & 3.06 & 2.67 & 2.34 & 2.13 \\
Therm-FM-T  & \textbf{8.04} & \textbf{1.71} & 1.04 & 0.87 & 0.71 & 0.72 \\
Therm-FM-B  & 12.42 & 2.10 & 1.24 & 0.96 & 0.81 & 0.62 \\
Therm-FM-L  & 14.66 & 2.48 & 1.02 & \textbf{0.85} & 0.72 & 0.67 \\
\bottomrule
\end{tabular}%
}
\end{minipage}\hfill
\begin{minipage}[t]{0.49\textwidth}
\centering
\textbf{(b) Pooled $R^2$; higher is better.}\par\vspace{2pt}
\resizebox{\linewidth}{!}{%
\begin{tabular}{lrrrrrr}
\toprule
Method & 0 & 10 & 50 & 100 & 250 & 500 \\
\midrule
U-FNO       & -0.2957 & 0.9889 & 0.9977 & 0.9985 & 0.9985 & \textbf{0.9993} \\
SAU-FNO     & -0.1794 & 0.9897 & 0.9974 & 0.9980 & \textbf{0.9990} & 0.9990 \\
FNO         & -1.7581 & 0.9875 & 0.9976 & 0.9983 & 0.9988 & 0.9990 \\
U-Net       & 0.3336 & 0.9938 & \textbf{0.9980} & 0.9983 & 0.9989 & 0.9990 \\
DeepOHeat   & 0.0361 & 0.9256 & 0.9784 & 0.9842 & 0.9883 & 0.9899 \\
Therm-FM-T  & \textbf{0.9056} & \textbf{0.9946} & 0.9979 & 0.9985 & 0.9989 & 0.9989 \\
Therm-FM-B  & 0.6738 & 0.9923 & 0.9971 & 0.9982 & 0.9987 & 0.9992 \\
Therm-FM-L  & 0.6698 & 0.9889 & \textbf{0.9980} & \textbf{0.9986} & \textbf{0.9990} & 0.9990 \\
\bottomrule
\end{tabular}%
}
\end{minipage}
\end{table*}

\begin{table*}[!t]
\caption{Complete target-case adaptation results for (a) per-sample MaxAE
and (b) peak-temperature error $E_{T_{\max}}$, both in kelvin.}
\label{tab:appendix-adaptation-max-peak}
\centering
\scriptsize
\begin{minipage}[t]{0.49\textwidth}
\centering
\textbf{(a) MaxAE; lower is better.}\par\vspace{2pt}
\resizebox{\linewidth}{!}{%
\begin{tabular}{lrrrrrr}
\toprule
Method & 0 & 10 & 50 & 100 & 250 & 500 \\
\midrule
U-FNO       & 39.50 & 8.68 & 5.94 & 5.21 & 4.92 & 4.47 \\
SAU-FNO     & 36.49 & 7.87 & 4.86 & 4.55 & \textbf{3.71} & \textbf{3.66} \\
FNO         & 66.03 & 10.91 & 5.59 & 4.71 & 4.02 & 3.82 \\
U-Net       & 32.68 &  9.42 &  7.40 & 6.85 & 5.97 & 5.54 \\
DeepOHeat   & 33.49 & 12.57 & 9.34 & 8.82 & 8.21 & 8.12 \\
Therm-FM-T  & \textbf{15.60} & \textbf{6.70} & 5.28 & 4.84 & 4.45 & 4.49 \\
Therm-FM-B  & 21.34 & 7.08 & 5.15 & 4.66 & 4.34 & 4.45 \\
Therm-FM-L  & 24.16 & 7.96 & \textbf{4.81} & \textbf{4.39} & 4.24 & 4.24 \\
\bottomrule
\end{tabular}%
}
\end{minipage}\hfill
\begin{minipage}[t]{0.49\textwidth}
\centering
\textbf{(b) $E_{T_{\max}}$; lower is better.}\par\vspace{2pt}
\resizebox{\linewidth}{!}{%
\begin{tabular}{lrrrrrr}
\toprule
Method & 0 & 10 & 50 & 100 & 250 & 500 \\
\midrule
U-FNO       & 31.68 & 3.07 & 1.33 & 1.01 & 0.94 & 0.82 \\
SAU-FNO     & 31.94 & 3.16 & 1.72 & 1.29 & 0.86 & 1.16 \\
FNO         & 60.10 & 3.03 & 1.47 & 1.35 & 0.87 & 0.88 \\
U-Net       & 24.41 & 2.48 & \textbf{1.23} & 1.06 & 0.93 & 0.92 \\
DeepOHeat   & 25.62 & 6.78 & 4.12 & 3.90 & 3.48 & 3.28 \\
Therm-FM-T  & \textbf{10.53} & \textbf{2.39} & 1.55 & 1.24 & 0.92 & 0.99 \\
Therm-FM-B  & 16.82 & 2.90 & 1.45 & 1.09 & \textbf{0.77} & 0.77 \\
Therm-FM-L  & 17.21 & 3.21 & 1.26 & \textbf{0.93} & 0.83 & \textbf{0.65} \\
\bottomrule
\end{tabular}%
}
\end{minipage}
\end{table*}

\begin{table*}[!t]
\caption{Complete Top-50 MAE (K) under target-case adaptation.  The metric is
computed at the 50 hottest ground-truth locations of each sample.}
\label{tab:appendix-adaptation-top50}
\centering
\setlength{\tabcolsep}{8.0pt}
\begin{tabular}{lrrrrrr}
\toprule
Method & $K=0$ & $K=10$ & $K=50$ & $K=100$ & $K=250$ & $K=500$ \\
\midrule
U-FNO       & 26.65 & 3.15 & 1.49 & 1.12 & 0.99 & 0.88 \\
SAU-FNO     & 26.99 & 3.32 & 1.89 & 1.45 & 0.95 & 1.20 \\
FNO         & 39.86 & 3.30 & 1.75 & 1.49 & 0.98 & 1.02 \\
U-Net       & 18.39 & \textbf{2.24} & \textbf{1.28} & \textbf{1.02} & 1.00 & 0.91 \\
DeepOHeat   & 24.08 & 6.76 & 4.61 & 4.21 & 3.75 & 3.59 \\
Therm-FM-T  & \textbf{9.08} & 2.37 & 1.64 & 1.37 & 1.10 & 1.08 \\
Therm-FM-B  & 15.77 & 3.05 & 1.71 & 1.40 & \textbf{0.91} & 0.95 \\
Therm-FM-L  & 13.78 & 2.96 & 1.45 & 1.04 & 0.98 & \textbf{0.86} \\
\bottomrule
\end{tabular}
\end{table*}

The complete tables reinforce two points hidden by RMSE alone.  First, the
best model changes with both label budget and metric: Therm-FM-T leads
all six metrics at $K=0$ and all but Top-50 MAE at $K=10$, where U-Net leads;
at $K=50$ the lead is split between Therm-FM-L (RMSE and MaxAE) and U-Net
(MAE, $E_{T_{\max}}$, and Top-50 MAE), with the two tied on $R^2$; at $K=100$
Therm-FM-L co-leads RMSE with U-FNO and leads every other metric except
Top-50 MAE (U-Net); and at
$K\ge250$ the operator baselines take over the global-field metrics (SAU-FNO
at $K=250$, tied with Therm-FM-L on $R^2$; U-FNO at $K=500$) and SAU-FNO the
worst-cell error, while the
hotspot-oriented metrics are led by Therm-FM-B at $K=250$ and Therm-FM-L at
$K=500$.  Second, the
first 10 labels per case recover most of the global correlation
($R^2\ge0.92$ for every model), but
worst-cell errors remain several kelvin and improve more slowly.

\begin{figure*}[!tbp]
  \centering
  \includegraphics[width=0.82\textwidth]{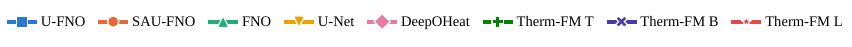}\\[2pt]
  \begin{minipage}[t]{0.158\textwidth}\centering
    \includegraphics[width=\linewidth]{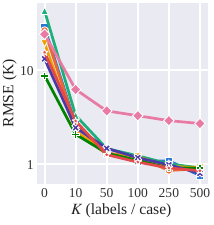}\\[1pt]
    {\footnotesize\bfseries (a) RMSE}
  \end{minipage}\hfill
  \begin{minipage}[t]{0.158\textwidth}\centering
    \includegraphics[width=\linewidth]{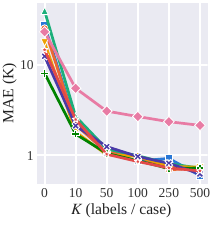}\\[1pt]
    {\footnotesize\bfseries (b) MAE}
  \end{minipage}\hfill
  \begin{minipage}[t]{0.158\textwidth}\centering
    \includegraphics[width=\linewidth]{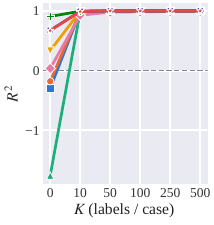}\\[1pt]
    {\footnotesize\bfseries (c) Pooled $R^2$}
  \end{minipage}\hfill
  \begin{minipage}[t]{0.158\textwidth}\centering
    \includegraphics[width=\linewidth]{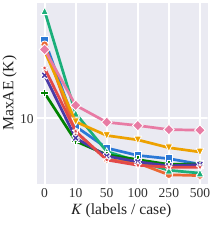}\\[1pt]
    {\footnotesize\bfseries (d) MaxAE}
  \end{minipage}\hfill
  \begin{minipage}[t]{0.158\textwidth}\centering
    \includegraphics[width=\linewidth]{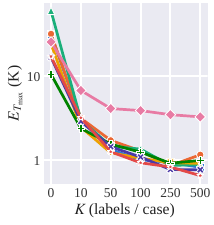}\\[1pt]
    {\footnotesize\bfseries (e) $E_{T_{\max}}$}
  \end{minipage}\hfill
  \begin{minipage}[t]{0.158\textwidth}\centering
    \includegraphics[width=\linewidth]{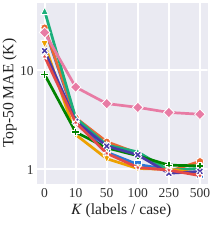}\\[1pt]
    {\footnotesize\bfseries (f) \topmae{}}
  \end{minipage}
  \caption{Target-case adaptation on \scopeid{5} for all six benchmark metrics.
  $K$ is the number of labeled samples per held-out case ($K=0$ is the frozen
  \scopeid{4} checkpoint on the common holdout); error metrics use a
  logarithmic axis. The steep improvement at the first few labeled samples is
  consistent across model families, while the remaining gaps depend on the
  error criterion.}
  \label{fig:appendix-fewshot-metrics}
\end{figure*}

\paragraph{Optimization record.}
All adaptation tables use the final checkpoint after the same 50-epoch
budget and evaluate every five epochs.  The release stores the intermediate
JSON records so that convergence and checkpoint-selection studies can be
reconstructed without changing the official final-checkpoint comparison.
The appendix freezes the complete final-checkpoint numbers for the eight
registered baselines; intermediate traces remain machine-readable rather
than being summarized by an additional, potentially redundant figure.

\section{Training Cost and Efficiency Details}
\label{app:training-cost}

Table~\ref{tab:appendix-training-cost} reports model-training cost under the
same released splits and hardware protocol.  Wall time measures the standard
training run, while GPU-s multiplies wall time by the number of GPUs and is
included only to expose aggregate accelerator occupancy.  Neither quantity
includes one-time HotSpot label generation, data conversion, checkpoint
download, or manual hyperparameter exploration.

\begin{table*}[!t]
\caption{Parameter count and training cost across \scoperange{2}{4}.
GPU-s is wall time multiplied by GPU count and should not be read as
hardware-normalized throughput. All methods use a single GPU, so wall time and GPU-s coincide.}
\label{tab:appendix-training-cost}
\centering
\setlength{\tabcolsep}{9.5pt}
\begin{tabular}{lrrrrrrr}
\toprule
& & \multicolumn{2}{c}{\scopeid{2}} & \multicolumn{2}{c}{\scopeid{3}} & \multicolumn{2}{c}{\scopeid{4}} \\
\cmidrule(lr){3-4}\cmidrule(lr){5-6}\cmidrule(lr){7-8}
Method & Params. & Wall (s) & GPU-s & Wall (s) & GPU-s & Wall (s) & GPU-s \\
\midrule
FNO        & 11.98M  &  1844 &  1844 &  1857 &  1857 &  1867 &  1867 \\
U-Net      & 5.06M   &   845 &   845 &   841 &   841 &   875 &   875 \\
U-FNO      & 5.10M   & 11603 & 11603 & 11606 & 11606 & 11658 & 11658 \\
SAU-FNO    & 5.11M   & 22193 & 22193 & 22299 & 22299 & 22300 & 22300 \\
DeepOHeat  & 7.92M   &   194 &   194 &   228 &   228 &   264 &   264 \\
Therm-FM-T & 20.7M   &  3195 &  3195 &  3245 &  3245 &  3297 &  3297 \\
Therm-FM-B & 157.6M  &  6283 &  6283 &  6330 &  6330 &  6580 &  6580 \\
Therm-FM-L & 628.4M  & 10782 & 10782 & 10806 & 10806 & 11030 & 11030 \\
\bottomrule
\end{tabular}
\end{table*}

The table is intended for reproducibility, not for ranking architecture
efficiency across different accelerators.

DeepOHeat has the shortest wall time but also the largest errors across
the S2--S4 within-family evaluations, so speed alone is not a sufficient selection criterion.  Among the
strong Fourier baselines, U-FNO is substantially cheaper than SAU-FNO under
the recorded recipe, and it is also more accurate on the global-field
errors of every within-family Scope; SAU-FNO's additional attention cost
mainly buys a lower worst-cell error at \scopeid{2} and \scopeid{4}.
Therm-FM-T has a modest wall time; the B and L variants increase
aggregate accelerator use by about $2.0\times$ and $3.3\times$ at
\scopeid{4}, improve within-family accuracy over T on every within-family
Scope (although L does not improve over B at \scopeid{3}), and transfer
worse in frozen OOD evaluation.  These comparisons
motivate reporting error, wall time, GPU count, and parameter scale together.

\paragraph{Cost-accounting boundary.}
The measured interval begins with the first optimizer update and ends after
the final training epoch.  It excludes one-time case generation, HotSpot
label production, CSV-to-tensor conversion, checkpoint download, and manual
hyperparameter exploration.  Label generation is excluded because it is a
dataset-production cost shared by all fully supervised models; for target
adaptation it becomes a deployment-specific cost and should be reported
separately.  This boundary makes the model-training numbers reproducible
without presenting them as end-to-end system latency.

\paragraph{Why wall time and GPU-s are both retained.}
Wall time approximates researcher turnaround on the recorded machine,
whereas GPU-s exposes aggregate occupancy.  Neither is hardware-normalized
throughput: four-GPU execution can reduce elapsed time while increasing total
accelerator use, and different kernels exploit the RTX A6000 differently.  A new
result should therefore state device model, GPU count, precision, software
stack, epoch count, batch size, and parameter count alongside both measures.
When hardware changes, the released commands and output schema remain
comparable even if raw timing does not.

\paragraph{Accuracy--cost selection.}
The table supports deployment choices rather than one universal efficiency
rank.  A fast screening loop may prefer U-Net despite lower accuracy; a
    Within-Family DSE study may justify the additional cost of U-FNO or SAU-FNO;
and a pretrained Therm-FM checkpoint may be attractive when \scopeid{1}
accuracy is central. Cross-Package OOD remains the decisive caveat: none of the
cost advantages makes the frozen \scopeid{5} errors operationally small, so
adaptation label and optimization costs must be included whenever few-shot
recovery is part of the intended workflow.

\section{Generated-Scope Visualizations}
\label{app:generated-viz}
\begin{figure*}[!tbp]
  \centering
  \includegraphics[width=0.72\textwidth]{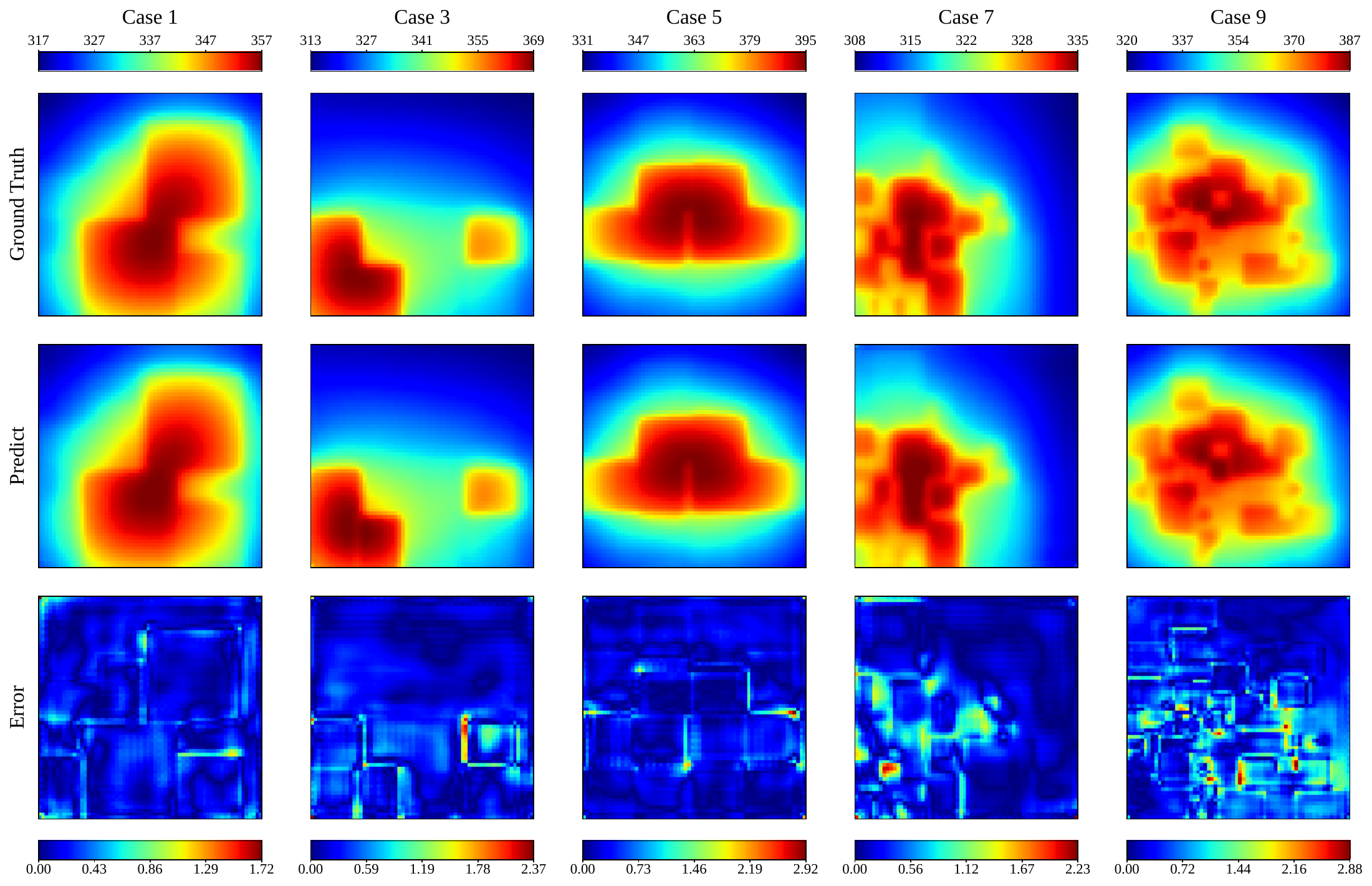}
  \caption{\scopeid{2} visualization with U-FNO. Columns correspond
  to Cases~1, 3, 5, 7, and~9, with the median-RMSE test sample shown
  for each case. Rows show ground truth, prediction, and absolute
  error in kelvin.}
  \label{fig:appendix-s2-visualization}
\end{figure*}

\begin{figure*}[!tbp]
  \centering
  \includegraphics[width=0.72\textwidth]{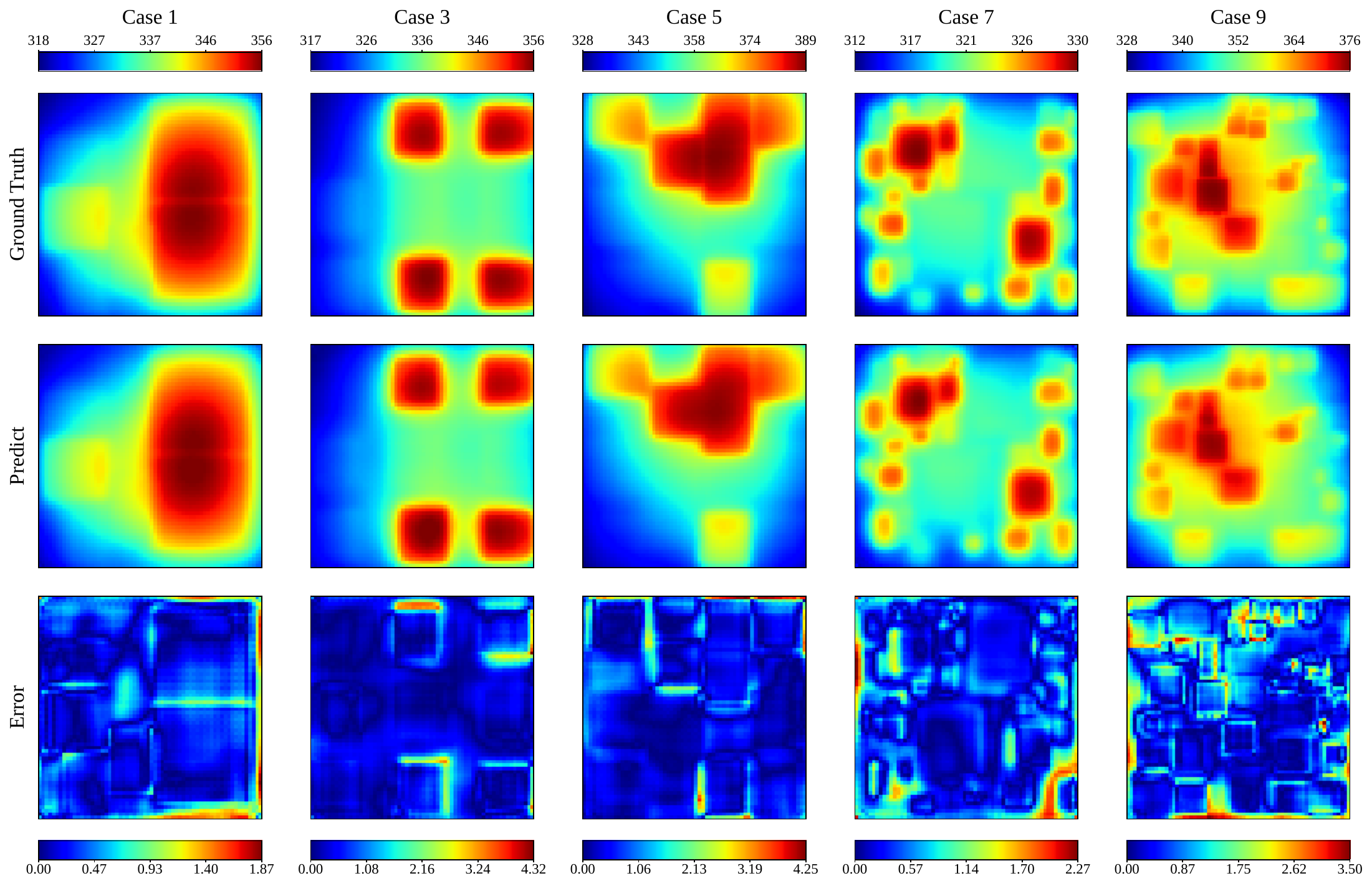}
  \caption{\scopeid{3} visualization with U-FNO. Columns correspond
  to Cases~1, 3, 5, 7, and~9, with the median-RMSE test sample shown
  for each case. Rows show ground truth, prediction, and absolute
  error in kelvin.}
  \label{fig:appendix-s3-visualization}
\end{figure*}

\begin{figure*}[!tbp]
  \centering
  \includegraphics[width=0.72\textwidth]{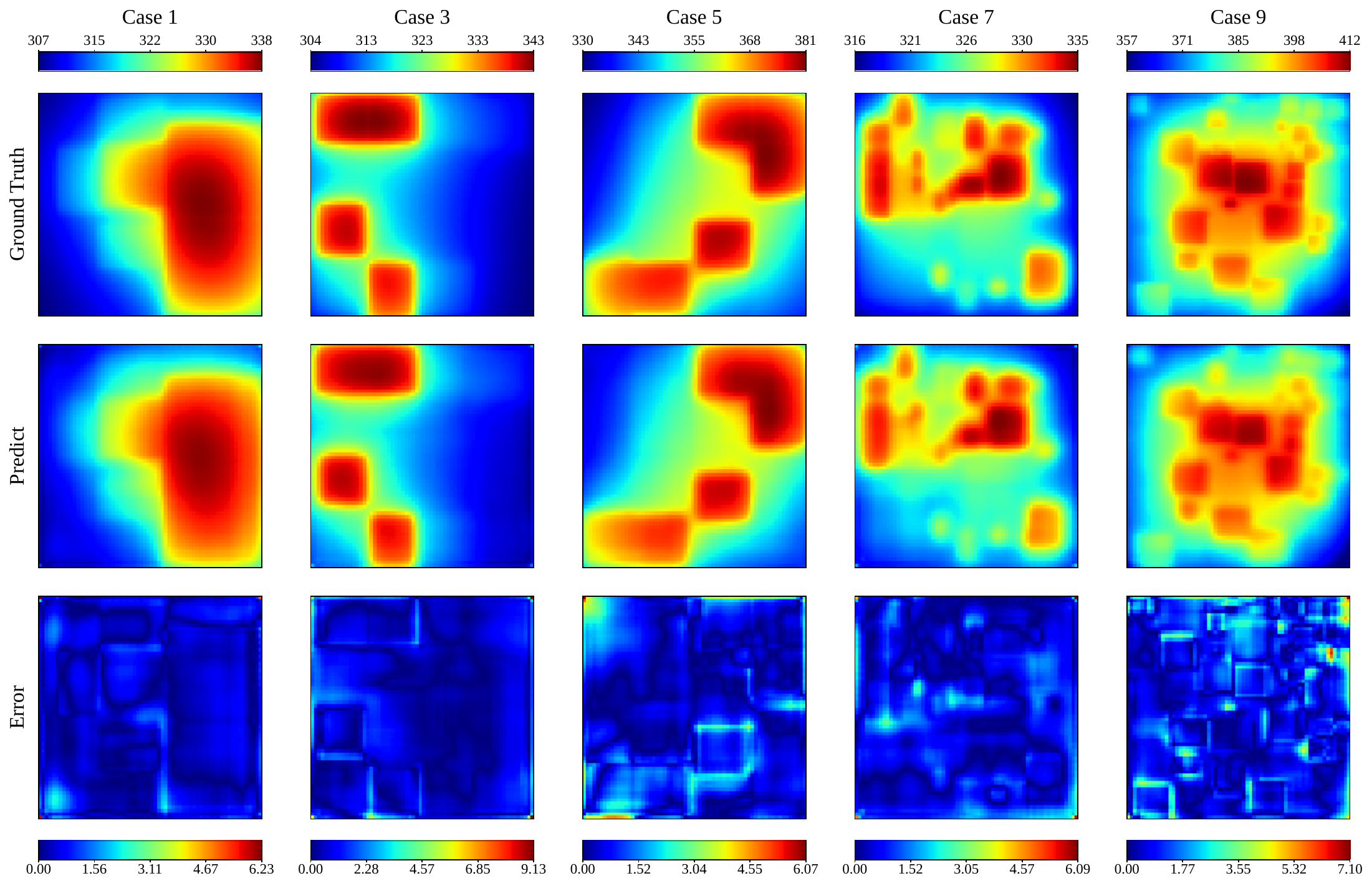}
  \caption{\scopeid{4} visualization with U-FNO. Columns correspond
  to Cases~1, 3, 5, 7, and~9, with the median-RMSE test sample shown
  for each case. Rows show ground truth, prediction, and absolute
  error in kelvin.}
  \label{fig:appendix-s4-visualization}
\end{figure*}

\begin{figure*}[!t]
  \centering
  \includegraphics[width=0.72\textwidth]{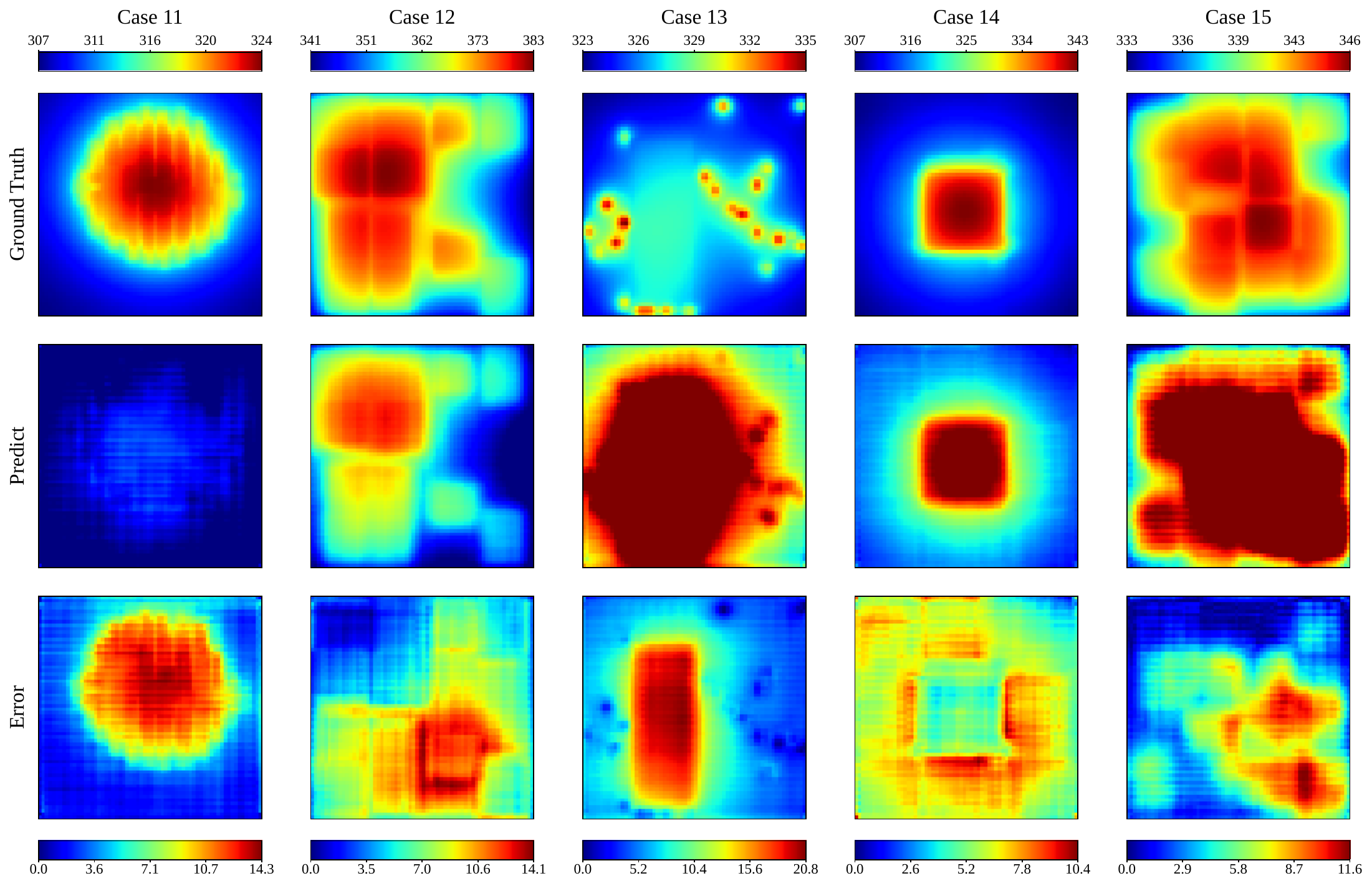}
  \caption{\scopeid{5} zero-shot visualization with Therm-FM-T, the
  lowest-RMSE frozen model, on the held-out Cases~11--15 without target labels
  (the median-RMSE sample of each case). The predictor misses, merges, or
  inflates hot regions in four of the five unseen packages, matching the
  structural failure quantified in Table~\ref{tab:s5-per-case-main}.}
  \label{fig:appendix-s5-visualization}
\end{figure*}

\begin{figure*}[!t]
  \centering
  \includegraphics[width=0.72\textwidth]{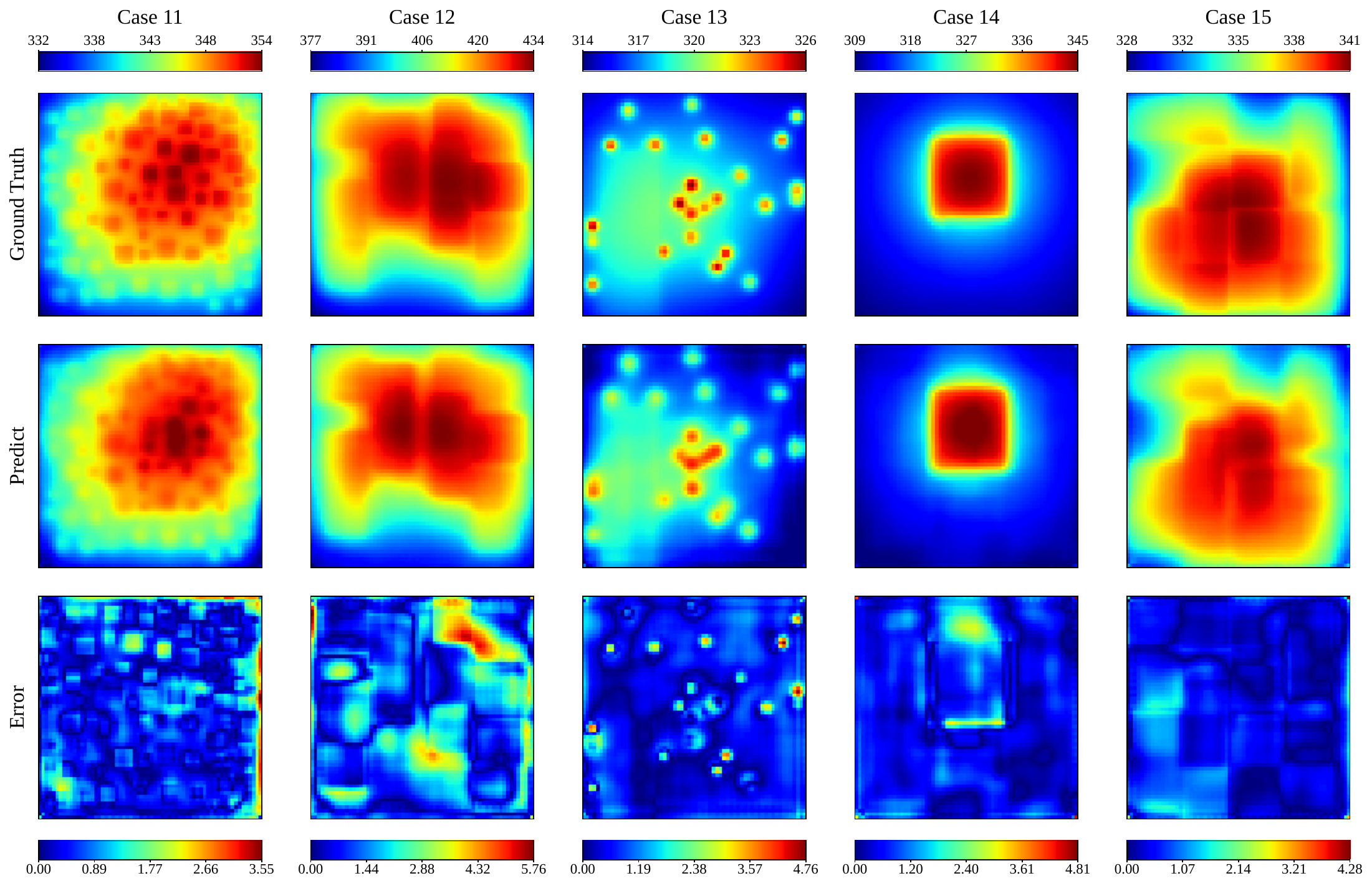}
  \caption{\scopeid{5} visualization with U-FNO after target-case
  adaptation using $K=100$ labeled samples per case. Columns correspond
  to Cases~11--15, with the median-RMSE sample of each case shown
  from the common 2,500-sample holdout. Rows show ground truth,
  prediction, and absolute error in kelvin. The adapted predictions
  recover the global temperature structure on all five cases, including
  the scattered micro-chiplet hot spots of Case~13.}
  \label{fig:appendix-s5-k100-visualization}
\end{figure*}

Figures~\ref{fig:appendix-s2-visualization}--\ref{fig:appendix-s5-k100-visualization}
show representative predictions for \scoperange{2}{5} using the models
specified in the corresponding captions, in the same format as the
\scopeid{1} visualizations
(Fig.~\ref{fig:appendix-l1-visualizations}). For \scoperange{2}{4},
columns correspond to Cases~1, 3, 5, 7, and~9 from each Scope's own
test split. For \scopeid{5}, columns correspond to Cases~11--15 under
zero-shot evaluation and after target-case adaptation using $K=100$
labeled samples per case. Each column shows the median-RMSE sample
within that case for the displayed model and evaluation setting.
Ground truth and prediction share the temperature scale above each
column, with out-of-range predictions clipped for visualization;
the absolute-error panel uses a separate scale below.

\paragraph{Reading \scoperange{2}{4}.}
Within the represented system families the predicted fields are visually
indistinguishable from the HotSpot ground truth at the scale of the figure.
The residual error is not spatially uniform: it concentrates along chiplet
boundaries, where the temperature gradient is steepest, and at the corners of
high-power dies, while the package interior and the background are
reproduced to a fraction of a kelvin. Adding the material channel
(\scopeid{3}) and the cooling channels (\scopeid{4}) raises the error scale
at the worst cell from roughly 1.7--2.9~K (\scopeid{2}) to 1.9--4.3~K
(\scopeid{3}) and 6.1--9.1~K (\scopeid{4}); at \scopeid{4} this worst cell
lies at a grid corner in every displayed sample. The error pattern
keeps the same edge-localized character, and the additional physics is absorbed
rather than breaking the predictor. These are the same median samples whose
Scope-level averages appear in Tables~\ref{tab:appendix-s2-full}--%
\ref{tab:appendix-s4-full}.

\paragraph{Reading \scopeid{5} zero-shot.}
Figure~\ref{fig:appendix-s5-visualization} shows the strongest frozen model,
Therm-FM-T, and it is still qualitatively wrong on four of the five
compound cross-package OOD cases.
The failure is structural rather than numerical: on Case~11 the hot
center of the 80-chiplet array is missed almost entirely and the field is
predicted near the bottom of the temperature range, on Case~12 only part of
the hot region is recovered, on Case~13 the scattered micro-chiplets are
merged into a single saturated hot plateau, and on Case~15 the stacked dies
collapse into one oversized hot block. Case~14 is the exception: the hot spot
of the dominant chiplet is predicted with the correct position and footprint,
but the heat spreading around it is overestimated, leaving a ring-shaped
error. The error maps reach 10--21~K and follow the misplaced structures
instead of the chiplet edges. This is the visual counterpart of the order-of-magnitude jump in
Table~\ref{tab:appendix-s5-all-metrics}: a predictor that is accurate across
diverse layouts, materials, and cooling conditions of the represented families
does not carry a usable thermal mapping to an unseen package structure, and no
baseline in the benchmark is exempt from this.

\paragraph{Reading \scopeid{5} after adaptation.}
Figure~\ref{fig:appendix-s5-k100-visualization} applies the same protocol to
U-FNO, the strongest operator baseline at $K=100$, after 100 labeled samples per OOD
case. With 500 target labels in total, the global thermal feature is recovered on
all five packages: the chiplet array of Case~11, the hot region of
Case~12, the dominant die of Case~14, and the stacked dies of Case~15 are
placed and scaled correctly, and the scattered micro-chiplets of Case~13
largely reappear as separate hot spots at the correct positions. The error
scale drops from 10--21~K to 3.5--5.8~K at the worst cell and to below one
kelvin over most of the field. The remaining residuals are again
edge-localized, approaching the error pattern observed in the S2--S4
within-family evaluations; Case~12, which reaches the highest
temperatures, retains the largest residual (1.93~K case-mean RMSE, versus
0.74--0.89~K on the other four cases).
Target-case adaptation is therefore a practical remedy for cross-package
OOD failure---limited target-domain labels can move an otherwise poor zero-shot
prediction substantially toward the S2--S4 accuracy regime---but this
recovery requires target labels and is reported separately from, not as a
substitute for, zero-shot cross-package generalization.


\section{Benchmark Extension Protocol}
\label{app:extension}

\benchname{} reduces extension effort through reusable data and model
interfaces.
New datasets compatible with the current tensor representation and
split convention can be supplied as input/output MAT pairs and selected
by dataset name, reusing the shared loading, normalization,
and evaluation routines.
Model integration follows the interface described in
Appendix~\ref{app:training-details}.

Extensions involving new simulators, measurements, or temporal
representations may require task-specific conversion, configuration,
or adapters, but can build on these shared interfaces.
Each extension should publish its case manifest, channel semantics,
label provenance, fixed split identifiers, and reference metrics,
and specify whether it broadens a Generalization Scope or defines
a separate domain-shift track, while preserving the primary benchmark
tracks.